\documentclass{WileyMSP-template}
\usepackage{amsmath}
\usepackage{amssymb}
\usepackage{mathtools}
\usepackage{xcolor}
\usepackage{bm}
\usepackage{appendix}

\begin{document}
\pagestyle{fancy}

\rhead{\includegraphics[width=2.5cm]{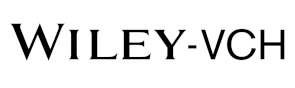}}


\title{Two-Time Quantum Fluctuations Approach and its Relation to the Bethe--Salpeter Equation}

\maketitle

\author{Erik Schroedter*}
\author{Michael Bonitz}

\dedication{}

\begin{affiliations}
Erik Schroedter, Michael Bonitz\\
Address: Institut f\"ur Theoretische Physik und Astrophysik, Christian-Albrechts-Universit\"at zu Kiel, 24098 Kiel, Germany\\
Email Address: schroedter@physik.uni-kiel.de

\end{affiliations}




\begin{abstract}
    Correlated quantum many-particle systems out of equilibrium are of high interest in many fields, including correlated solids, ultracold atoms or dense plasmas. Accurate theoretical description of these systems is challenging both, conceptionally and with respect to computational resources. We have recently presented a quantum fluctuations approach which is equivalent to the nonequilibrium $GW$ approximation [E. Schroedter \textit{et al.}, Cond. Matt. Phys. \textbf{25}, 23401 (2022)] that promises high accuracy at low computational cost. The method exhibits CPU time scaling that is linear in the number of time steps, like the G1--G2 scheme [Schlünzen \textit{et al.}, Phys. Rev. Lett. \textbf{124}, 076601 (2020)], however, at a much reduced computer memory cost. In a second publication [E. Schroedter \textit{et al.}, Phys. Rev. B \textbf{108}, 205109 (2023)], this approach was extended to the two-time exchange-correlation functions and the dynamic density response properties. Here, we analyze the properties of this approach in more detail. We establish the physical meaning of the central approximation -- the quantum polarization approximation. We demonstrate that the method is equivalent to the Bethe--Salpeter equation for the two-time exchange-correlation function when the generalized Kadanoff-Baym ansatz with Hartree-Fock propagators is applied.
\end{abstract}


\section{Introduction}\label{s:intro}
The study of the behavior of quantum many-body systems  following an external excitation is a topic of great interest across various fields, such as dense plasmas, nuclear matter, ultracold atoms, and correlated solids. Multiple methods are available for simulating such systems, including real-time quantum Monte Carlo, density matrix renormalization group techniques, time-dependent density functional theory, and quantum kinetic theory. In experimental investigations of many-particle systems, correlation functions of density or spin fluctuations and their corresponding dynamic structure factors play a central role, see e.g. Ref. \cite{giuliani_vignale_2005} for an overview. To accurately compute these quantities, taking correlation effects into consideration, various equilibrium simulations are employed. For correlated solids and warm dense matter, the most precise results are often obtained through quantum Monte Carlo simulations \cite{moreo_prb_93,lee_prb_03,assaad_prb_06,dornheim_physrep_18,dornheim_prl_18,dornheim_pop_23,bonitz_pop_20}, which also include the analysis of non-linear responses \cite{bonitz-etal.94pre,dornheim_prl_20,dornheim_prr_21}.

In addition to equilibrium simulations, a range of nonequilibrium approaches is available, such as dynamical mean field theory (DMFT), e.g., \cite{gull_prb_19,gull_prr_20}, time-dependent density matrix renormalization group (DMRG), e.g., \cite{pereira_prb_12}, and nonequilibrium Green functions (NEGF), cf.  \cite{kwong_prl_00} and references therein.
Here, we focus on the NEGF approach \cite{keldysh64,stefanucci-book,balzer-book,bonitz_pss_19_keldysh}, which offers a rigorous description of the quantum dynamics of correlated systems in multiple dimensions, e.g., \cite{schluenzen_prb16}. However, NEGF simulations are computationally expensive, primarily due to their cubic scaling with simulation time $N_t$ (number of time steps). Recently linear scaling with $N_t$ has become feasible within the G1--G2 scheme \cite{schluenzen_prl_20,joost_prb_20}, which could be demonstrated even for advanced selfenergy approximations like $GW$ and the $T$-matrix. Additionally, the nonequilibrium dynamically screened ladder approximation is also becoming viable, at least for lattice models \cite{joost_prb_22,donsa_prr_23}, for details of the scheme, see Ref.~\cite{bonitz_pssb23}.
The advantage of the linear scaling in the G1--G2 scheme comes with a cost: the simultaneous propagation of the time-diagonal single-particle and correlated two-particle Green functions, $G_1(t)$ and $\mathcal{G}_2(t)$, demands a substantial computational effort for computing and storing all matrix elements of $\mathcal{G}_2$. For instance, the CPU time of $GW$-G1--G2 simulations scales with $N_b^6$, where $N_b$ denotes the size of the basis. Moreover, going to a time-local formulation of the many-body problem severely restricts access to aforementioned correlation functions as they, for nonequilibrium systems, depend on multiple times. Thus, it is essential to explore alternative formulations that are more suitable for computational purposes, ideally without sacrificing accuracy.

In Ref. \cite{schroedter_cmp_22}, the present authors introduced an alternative formulation of the quantum many-body problem that is  based on a stochastic approach to the  dynamics of quantum fluctuations. Building upon earlier stochastic concepts in classical  kinetic theory by Klimontovich, e.g., \cite{klimontovich_jetp_57,klimontovich_jetp_72,klimontovich_1982}, and the work of Ayik, Lacroix~\cite{ayik_plb_08,lacroix_prb14,lacroix_epj_14}, and others, e.g., \cite{filinov_prb_2,polkovnikov_ap_10}, on stochastic approaches to describing the dynamics of quantum systems, an equation of motion for single-particle fluctuations, $\delta \hat{G}$, was derived that constitutes the basis of the quantum polarization approximation (PA) and that was shown to be closely related to the nonequilibrium $GW$ approximation within the G1--G2 scheme. More specifically, it was shown that the PA and the $GW$ approximation with additional exchange contributions, which will be denoted with $GW^\pm$ in the following, are equivalent in the weak coupling limit.
An advantage of the quantum fluctuations approach is that it allows for a straightforward extension from a time-local to a two-time description of the many-body system. Furthermore, in Ref.~\cite{schroedter_23}, an extension of the stochastic approach to the so-called multiple ensembles (ME) approach was presented, which allows for the computation of commutators of operators and, thus, density response functions and their dynamic structure factors, both in the ground state and for systems far from equilibrium following an external excitation. Most importantly, this extension is applicable to large systems and long simulation times. However, in Ref.~\cite{schroedter_23}, the physical meaning of the quantum PA has been left open, and its equivalence to the Bethe-Salpeter equation of NEGF theory was not demonstrated.

In this work, we close this gap and extend the results of Refs. \cite{schroedter_cmp_22,schroedter_23} and show that the equivalence of the quantum polarization approximation and the nonequilibrium $GW^\pm$ approximation extends from the time-local case to the two-time case. To this end, we consider the relation of the quantum fluctuations approach in its two-time formulation to the Bethe--Salpeter equation and derive a two-time generalization of the correlated two-particle Green function, $\mathcal{G}_2(t,t')$, as well as its equations of motion. 

This paper is structured as follows. In Sec.~\ref{s:theoretical_framework}, we set up the necessary theoretical framework and introduce nonequilibrium Green functions and the exchange-correlation function as well as their real-time components and the relation to the quantum fluctuations approach. Next, in Sec.~\ref{s:dynamics}, we discuss the dynamics of quantum many-body systems. More specifically, we consider the Keldysh--Kadanoff--Baym equations in their contour-time formulation as well as on the time diagonal and the equations of motion within the quantum fluctuations approach. In Sec.~\ref{s:QPA}, we discuss the two-time quantum polarization approximation and follow this up in Sec.~\ref{s:GWA} by the discussion of the two-time $GW$ approximation. Here, this is done by first considering the general Bethe--Salpeter approach and then considering the formulation of the $GW$ approximation within this framework. Moreover, we introduce the generalized Kadanoff--Baym ansatz and use it to derive the equations of motion for the two-time $GW$ approximation. In Sec.~\ref{s:equivalence}, we compare the two approximations, followed by an application to small Hubbard chains in Sec.~\ref{s:application}. Finally, we provide a summary and outlook in Sec.~\ref{s:discussion}.
\section{Theoretical Framework} \label{s:theoretical_framework}
\subsection{Nonequilibrium Green Functions and Exchange-Correlation Function}
The central quantity that provides the foundation for all  considerations below is the $n$-particle nonequilibrium Green function (NEGF)
\begin{align}\label{eq:def_n-particle_NEGF}
    G^{(n)}_{i_1\dots i_n j_1\dots j_n}(z_1,\dots,z_n,z'_1,\dots, z'_n)\coloneqq \Big(\frac{1}{\mathrm{i}\hbar}\Big)^n \Big\langle \mathcal{T}_\mathcal{C}\Big\{\hat{c}_{i_1}(z_1)\dots \hat{c}_{i_n}(z_n)\hat{c}^\dagger_{j_n}(z'_n)\dots \hat{c}^\dagger_{j_1}(z'_1)\Big\}\Big\rangle\,,
\end{align}
where $\hat{c}^\dagger_i(z)$ and $\hat{c}_i(z)$  denote the creation and annihilation operators, respectively, of particles in the single-particle orbital $\psi_i$ for contour times $z$ on the Keldysh contour $\mathcal{C}$. Further, the single-particle orbitals $(\psi_i)_i$ are assumed to provide an orthonormal basis of the underlying single-particle Hilbert space. Lastly, $\mathcal{T}_\mathcal{C}$ denotes the time-ordering operator on the Keldysh contour\footnote{Here, we use the convention that the relative ordering of operators does not change for equal times.}. \\
For the two-particle NEGF, we consider the following decomposition into the Hartree--Fock Green function $G^{(2),\mathrm{HF}}$ and its correlated part $\mathcal{G}$,
\begin{align}
    G^{(2)}_{ijkl}(z_1,z_2,z'_1,z'_2)\equiv G^{(2),\mathrm{HF}}_{ijkl}(z_1,z_2,z'_1,z'_2)+\mathcal{G}_{ijkl}(z_1,z_2,z'_1,z'_2)\,, \label{eq:decomposition_2pNEGF}
\end{align}
where the Hartree--Fock contribution itself can be decomposed into the Hartree and the Fock Green function which are denoted as $G^{(2),\mathrm{H}}$ and $G^{(2),\mathrm{F}}$, respectively. These contributions are expressed in terms of the single-particle NEGF, $G\coloneqq G^{(1)}$, as\footnote{Throughout this work, the upper (lower) sign is used for bosons (fermions).} 
\begin{align}
    G_{ijkl}^{(2),\mathrm{H}}(z_1,z_2,z'_1,z'_2)&\coloneqq G_{ik}(z_1,z'_1)G_{jl}(z_2,z'_2)\,,\label{eq:def_Hatree_NEGF}\\
    G_{ijkl}^{(2),\mathrm{F}}(z_1,z_2,z'_1,z'_2)&\coloneqq \pm G_{il}(z_1,z'_2)G_{jk}(z_2,z'_1)\,. \label{eq:def_Fock_NEGF}
\end{align}
Alternatively, we can consider a different decomposition of the two-particle NEGF by introducing the two-particle exchange-correlation (XC) function, defined as 
\begin{equation}
    L_{ijkl}(z_1,z_2,z'_1,z'_2)\coloneqq G^{(2),\mathrm{F}}_{ijkl}(z_1,z_2,z'_1,z'_2)+\mathcal{G}_{ijkl}(z_1,z_2,z'_1,z'_2)\,, \label{eq:def_XC_function}
\end{equation}
that combines all exchange and correlation contributions to the two-particle NEGF. 
\subsection{Real-Time Components of the NEGF}
\begin{figure}
    \centering
    \includegraphics[width=0.6\textwidth]{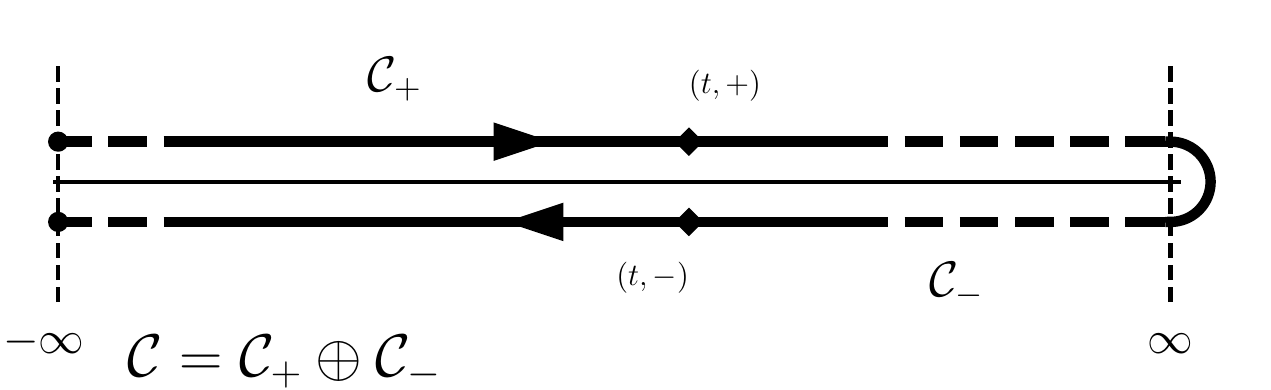}
    \caption{Illustration of the Keldysh contour \cite{keldysh64}, extending from $-\infty$ to $\infty$ and then in the reverse direction.}
    \label{fig:contour}
\end{figure}

In the following, we will consider the Keldysh contour shown in Fig.~\ref{fig:contour}, which is given by a combination of the causal branch, $\mathcal{C}_+$, extending from $-\infty$ to $\infty$ and the anti-causal branch, $\mathcal{C}_-$, going from $\infty$ to $-\infty$, i.e., we have
\begin{equation}
    \mathcal{C}=\mathcal{C}_+\oplus\mathcal{C}_-\,. \label{eq:contour}
\end{equation}
We can then express quantities on the Keldysh contour depending on a contour time $z$ in terms of a real time $t$ and an index $\alpha$ indicating the branch $z$ is on, i.e., the single-particle NEGF can be equivalently expressed as
\begin{equation}
    G_{ij}(z,z')\equiv G^{\alpha\alpha'}_{ij}(t,t')\,,
\end{equation}
where $\alpha,\alpha'\in \{+,-\}$. Consequently, we have four different real-time components of the single-particle NEGF\footnote{The Keldysh contour is sometimes defined denoting the causal branch as $\mathcal{C}_-$ and the anti-causal branch as $\mathcal{C}_+$ (see, for example, Ref.~\cite{stefanucci-book}). Consequently, the definitions of the real-time components change accordingly.} given by
\begin{align}
    G^\mathrm{c}_{ij}(t,t')&\coloneqq G^{++}_{ij}(t,t')= \frac{1}{\mathrm{i}\hbar} \Big\langle\mathcal{T}\Big\{\hat{c}_i(t)\hat{c}^\dagger_j(t')\Big\}\Big\rangle\,, \label{eq:def_causal_1pNEGF}\\
    G^<_{ij}(t,t')&\coloneqq G^{+-}_{ij}(t,t')= \pm \frac{1}{\mathrm{i}\hbar} \Big\langle \hat{c}^\dagger_j(t')\hat{c}_i(t)\Big\rangle\,,\label{eq:def_lesser_1pNEGF}\\
    G^>_{ij}(t,t')&\coloneqq G^{-+}_{ij}(t,t')= \frac{1}{\mathrm{i}\hbar} \Big\langle \hat{c}_i(t)\hat{c}^\dagger_j(t')\Big\rangle \,,\label{eq:def_greater_1pNEGF}\\
    G^\mathrm{a}_{ij}(t,t') &\coloneqq G^{--}_{ij}(t,t') = \frac{1}{\mathrm{i}\hbar}\Big\langle \Bar{\mathcal{T}}\Big\{ \hat{c}_i(t)\hat{c}^\dagger_j(t')\Big\}\Big\rangle\,, \label{eq:def_anticausal_1pNEGF}
\end{align}
where $\mathcal{T}$ denotes the chronological and $\Bar{\mathcal{T}}$ the anti-chronological time-ordering operator. The causal Green function $G^\mathrm{c}$ and the anti-causal Green function $G^\mathrm{a}$ can be expressed in terms of the greater ($G^>$) and lesser ($G^<$) component, i.e., 
\begin{align}
    G^\mathrm{c}_{ij}(t,t')&= \Theta_0(t,t')G^>_{ij}(t,t')+\Theta(t',t) G^<_{ij}(t,t') \,,\label{eq:G^c_G>/<}\\
    G^\mathrm{a}_{ij}(t,t')&= \Theta_0(t',t)G^>_{ij}(t,t')+\Theta(t,t') G^<_{ij}(t,t')\,, \label{eq:G^a_G>/<}
\end{align}
where $\Theta_0(t,t')\coloneqq1$, for $t\geq t'$, and $\Theta_0(t,t')\coloneqq0$, for $t<t'$, whereas  $\Theta$ is defined as $\Theta(t,t')\coloneqq1$, for $t>t'$, and $\Theta(t,t')\coloneqq0$, for $t\leq t'$. Further, we define two additional real-time components of the NEGF: the retarded and the advanced Green functions  given by
\begin{align}
    G^\mathrm{R}_{ij}(t,t')&\coloneqq \Theta_0(t,t')\Big\{G^>_{ij}(t,t')-G^<_{ij}(t,t')\Big\}\,,\label{eq:def_retarded_NEGF}\\
    G^\mathrm{A}_{ij}(t,t')&\coloneqq -\Theta_0(t',t)\Big\{ G^>_{ij}(t,t')-G^<_{ij}(t,t')\Big\}\,.\label{eq:def_advanced_NEGF}
\end{align}
On the time diagonal, we write for the lesser and greater components
\begin{equation}
    G^\gtrless_{ij}(t)\equiv G^\gtrless_{ij}(t,t)\,, 
\end{equation}
where both functions are related by
\begin{equation}
    G^>_{ij}(t)-G^<_{ij}(t)=\frac{1}{\mathrm{i}\hbar}\delta_{ij}\,. \label{eq:G</>_property}
\end{equation}
\subsection{Quantum Fluctuations}
Besides the real-time components of the single-particle NEGF, we define single-particle fluctuation operators on the time diagonal as 
\begin{equation}
    \delta\hat{G}_{ij}(t)\coloneqq \hat{G}^<_{ij}(t)-G^<_{ij}(t)\equiv \hat{G}^>_{ij}(t)-G^>_{ij}(t)\,, \label{eq:def_1pFluctuations}
\end{equation}
where we introduced the operator versions for the lesser and greater single-particle NEGF given by
\begin{align}
    \hat{G}^<_{ij}(t)&\coloneqq  \pm\frac{1}{\mathrm{i}\hbar}\hat{c}^\dagger_j(t)\hat{c}_i(t)\,,\label{eq:def_G<_op}\\
    \hat{G}^>_{ij}(t)&\coloneqq \frac{1}{\mathrm{i}\hbar}\hat{c}_i(t)\hat{c}_j^\dagger(t)\,. \label{eq:def_G>_op}
\end{align}
Notice that, due to the (anti)commutation relation of the creation and annihilation operators\footnote{The (anti)commutation relations are given by $[\hat{c}_i,\hat{c}_j^\dagger]_\mp =\delta_{ij}$ and $[\hat{c}^{(\dagger)}_i,\hat{c}^{(\dagger)}_j]_\mp=0$.}, the fluctuations of the greater and lesser NEGF coincide on the time diagonal. The single-particle fluctuation operators constitute the cornerstone of the quantum fluctuations approach as developed in Refs.~\cite{schroedter_cmp_22,schroedter_23}. Further, we define the correlation function of two single-particle fluctuations, denoted in the following as two-time two-particle fluctuations, as 
\begin{equation}
    L_{ijkl}(t,t')\coloneqq \Big\langle \delta\hat{G}_{ik}(t)\delta\hat{G}_{jl}(t')\Big\rangle\,. \label{eq:def_2pFluctuations}
\end{equation}
Here, we also denote two-particle fluctuations with $L$ as they are a special real-time component of the XC function, i.e., we have 
\begin{align}
    L^{-+-+}_{ijkl}(t,t',t,t')&= G^{(2),-+-+}_{ijkl}(t,t',t,t')- G^{--}_{ik}(t,t)G^{++}_{jl}(t',t')= L_{ijkl}(t,t')\,. \label{eq:XC_function-2pFluctuations}
\end{align}
Within the framework of the G1--G2 scheme \cite{schluenzen_prl_20,joost_prb_20,joost_prb_22}, the usually considered real-time component of the two-particle NEGF on the time diagonal is given by
\begin{equation}
    G^{(2),++--}_{ijkl}(t,t^+,t^-,t)=-\frac{1}{\hbar^2}\Big\langle \hat{c}^\dagger_k(t)\hat{c}^\dagger_l(t)\hat{c}_j(t)\hat{c}_i(t)\Big\rangle \,,
\end{equation}
as it corresponds to the two-particle reduced density matrix\footnote{Here, $t^{+/-}$ denotes the right/left-sided limit, i.e., $f(t^{+/-})\equiv\lim_{\epsilon \rightarrow 0^{+/-}} f(t+ \epsilon)$.}. The correlated part of this real-time component is usually also denoted as $\mathcal{G}$ and will be called two-particle correlations in the following. Moreover, it follows immediately from the (anti)commutation relations of the creation and annihilation operators that the following relation between the single-time two-particle fluctuations, $L(t)\coloneqq L(t,t)$, and two-particle correlations holds
\begin{align}
    L_{ijkl}(t)=\pm G^>_{il}(t)G^<_{jk}(t)+\mathcal{G}_{ijkl}(t)\,, \label{eq:L_G2_time-diagonal}
\end{align}
which implies that we have, for the correlated part,
\begin{equation}
    \mathcal{G}_{ijkl}(t)= \mathcal{G}_{ijkl}^{++--}(t,t^+,t^-,t)= \mathcal{G}_{ijkl}^{-+-+}(t,t,t,t)\,. \label{eq:relation_G2_real-time_components}
\end{equation}
The two-time extension of the two-particle correlation function, which will be considered in the following, is given by 
\begin{equation}
    \mathcal{G}_{ijkl}(t,t')\coloneqq \mathcal{G}^{-+-+}_{ijkl}(t,t',t,t') \,. \label{eq:def_2time-G2}
\end{equation}
Obviously, it holds that $\mathcal{G}(t,t)=\mathcal{G}(t)$, due to Eq.~\eqref{eq:relation_G2_real-time_components}, and 
\begin{equation}
    L_{ijkl}(t,t')=\pm G^>_{il}(t,t') G^<_{jk}(t',t) +\mathcal{G}_{ijkl}(t,t')\,, \label{eq:L_G2}
\end{equation}
due to the definition of the XC function, Eq.\eqref{eq:def_XC_function}. Further, we define two-particle ``source fluctuations'' as
\begin{equation}
    L_{ijkl}^{(0)}(t,t')\coloneqq \pm G^>_{il}(t,t') G^<_{jk}(t',t)\,,\label{eq:def_source_fluctuations}
\end{equation}
which are always present, even in the absence of pair interactions.

\section{Dynamics of Quantum Many-Body Systems} \label{s:dynamics}
\subsection{Keldysh--Kadanoff-Baym Equations} \label{ss:KBE}
We consider a generic Hamiltonian given by\footnote{We use Einstein's summation convention throughout this work.} 
\begin{equation}
    \hat{H}(t)= h_{ij}(t)\hat{c}^\dagger_i\hat{c}_j+\frac{1}{2}w_{ijkl}(t)\hat{c}^\dagger_i\hat{c}^\dagger_j\hat{c}_l\hat{c}_k\,, \label{eq:Hamiltonian}
\end{equation}
which includes a single-particle contribution $h$ and a pair interaction $w$. These two quantities are allowed to be time-dependent to describe, for example, time-dependent external fields or interaction quenches, respectively. The equations of motion (EOMs) for the single-particle two-time NEGF are the Keldysh--Kadanoff--Baym equations (KBE)
\begin{align}
    \mathrm{i}\hbar \partial_z G_{ij}(z,z')&=\delta_{ij}\delta_\mathcal{C}(z,z')+h_{ik}(z)G_{kj}(z,z')\pm \mathrm{i}\hbar w_{iklp}(z) G^{(2)}_{lpjk}(z,z,z',z^+)\,, \label{eq:KBE_1}\\
    -\mathrm{i}\hbar \partial_{z'}G_{ij}(z,z')&=\delta_{ij}\delta_\mathcal{C}(z,z')+G_{ik}(z,z')h_{kj}(z')\pm \mathrm{i}\hbar G^{(2)}_{iklp}(z,z'^-,z',z')w_{lpjk}(z')\,, \label{eq:KBE_2}
\end{align}
where $\delta_\mathcal{C}$ denotes the delta distribution on the Keldysh contour. \\
On the time diagonal, the EOM for the real-time component $G^<(t)$ has the following form\footnote{We define the commutator of matrix-valued quantities $A,B$ as $[A,B]_{ij}\coloneqq A_{ik}B_{kj}-B_{ik}A_{kj}$.}
\begin{align}
    \mathrm{i}\hbar \frac{\mathrm{d}}{\mathrm{d}t}G^<_{ij}(t)= \Big[h^\mathrm{HF},G^<\Big]_{ij}(t)+\Big[I+I^\dagger\Big]_{ij}(t)\,,\label{eq:time-diagonal_KBE}
\end{align}
where we introduced the effective single-particle Hartree--Fock Hamiltonian defined as 
\begin{equation}
    h^\mathrm{HF}_{ij}(t)\coloneqq h_{ij}(t)\pm \mathrm{i}\hbar w^\pm_{ikjl}(t)G^<_{lk}(t)\,,
\end{equation}
with the (anti)symmetrized pair interaction, $w^\pm_{ijkl}(t)\coloneqq w_{ijkl}(t)\pm w_{ijlk}(t)$. Moreover, we introduced the collision term defined as 
\begin{equation}
    I_{ij}(t)\coloneqq \pm\mathrm{i}\hbar w_{iklp}(t)\mathcal{G}_{plkj}(t)\,.
\end{equation}
Equivalently, we can rewrite Eq.~\eqref{eq:time-diagonal_KBE} using two-particle fluctuations instead of correlations, i.e., 
\begin{equation}
    \mathrm{i}\hbar\frac{\mathrm{d}}{\mathrm{d}t}G^<_{ij}(t)= \Big[h^\mathrm{H},G^<\Big]_{ij}(t)+\Big[\mathcal{I}+\mathcal{I}^\dagger\Big]_{ij}(t)\,,\label{eq:time-diagonal_KBE_L}
\end{equation}
where we have the effective single-particle Hartree Hamiltonian given by
\begin{equation}
    h_{ij}^\mathrm{H}(t)\coloneqq h_{ij}(t)\pm\mathrm{i}\hbar w_{ikjl}(t)G_{lk}(t)\,,
\end{equation}
and the collision term that is re-expressed in terms  of fluctuations:
\begin{equation}
    \mathcal{I}_{ij}(t)\coloneqq \pm\mathrm{i}\hbar w_{iklp}(t)L_{plkj}(t)\,. \label{eq:fluc_collision_term}
\end{equation}
\subsection{Dynamics of Fluctuations} \label{ss:dynamics_fluctuations}
As was shown in Ref.~\cite{schroedter_cmp_22}, the dynamics of quantum many-body systems can be described using the EOM for the lesser NEGF, Eq.~\eqref{eq:time-diagonal_KBE}, and the EOM for the single-particle fluctuation operators since the latter provides the basis of the EOMs of all $n$-particle fluctuations. The single-particle fluctuation operator obeys the following EOM
\begin{align}
    \mathrm{i}\hbar\frac{\mathrm{d}}{\mathrm{d}t}\delta\hat{G}_{ij}(t)=& \Big[h^\mathrm{H},\delta\hat{G}\Big]_{ij}(t)+\Big[\delta\hat{\Sigma}^\mathrm{H},G^<\Big]_{ij}(t)+\Big[\delta\hat{\mathcal{I}}+\delta\hat{\mathcal{I}}^\dagger\Big]_{ij}(t)\,, \label{eq:EOM_1pFluctuations}
\end{align}
where we introduced the fluctuations Hartree selfenergy operator defined as
\begin{equation}
    \delta\hat{\Sigma}^\mathrm{H}_{ij}(t)\coloneqq \pm\mathrm{i}\hbar w_{ikjl}(t) \delta\hat{G}_{lk}(t)\,,\label{eq:def_fluctuations_Hartree_selfenergy}
\end{equation}
and the second-order fluctuations collision term given by 
\begin{equation}
    \delta\hat{\mathcal{I}}_{ij}(t)\coloneqq \pm\mathrm{i}\hbar w_{iklp}(t)\delta\hat{L}_{plkj}(t)\,. \label{eq:def_2nd-order_collision_term}
\end{equation}
Here, we have introduced second-order fluctuation operators, i.e., fluctuations of fluctuations, defined as 
\begin{align}
    \delta\hat{L}_{ijkl}(t)\coloneqq \delta\hat{G}_{ik}(t)\delta\hat{G}_{jk}(t)-L_{ijkl}(t)\,.\label{eq:def_2nd-order_fluctuations}
\end{align}
The second-order fluctuations, $\delta\hat{\mathcal{I}}(t)$, give rise to a hierarchy of fluctuation equations that is similar to the BBGKY hierarchy of reduced density operators \cite{bonitz_qkt}. Here, the EOM for two-particle fluctuations couples to three-particle fluctuations and so forth \cite{schroedter_cmp_22}. As in the case of the BBGKY hierarchy, it is necessary to find decoupling approximations. Below, we will discuss the quantum polarization approximation, which is directly connected to the $GW$ approximation of many-particle theory.
\section{Two-Time Quantum Polarization Approximation} \label{s:QPA}
The quantum polarization approximation (PA) follows on the level of single-particle fluctuations by assuming 
\begin{equation}
    \delta\hat{L}_{ijkl}(t)\approx \delta\hat{L}^{(0)}_{ijkl}(t)\,,\label{eq:PA_assumption}
\end{equation}
where $\delta \hat{L}^{(0)}$ denotes the fluctuations of two-particle source fluctuations, Eq.~(\ref{eq:def_source_fluctuations}), given by 
\begin{equation}
    \delta\hat{L}^{(0)}_{ijkl}(t)\coloneqq \pm\Big\{G^>_{il}(t)\delta\hat{G}_{jk}(t)+\delta\hat{G}_{il}(t)G^<_{jk}(t)\Big\}\,. \label{eq:def_2p_source_fluctuations}
\end{equation}
This leads to the following EOM for single-particle fluctuation operators\footnote{Unless it is clear from the context or if an expression holds independently of the used approximation, we will indicate the framework in which an equation holds by using superscripts, e.g., $L^\mathrm{P}$ instead of $L$.}
\begin{align}
    \mathrm{i}\hbar\frac{\mathrm{d}}{\mathrm{d}t}\delta\hat{G}_{ij}(t)=&\Big[h^\mathrm{HF},\delta\hat{G}\Big]_{ij}(t)+\Big[\delta\hat{\Sigma}^\mathrm{HF},G^<  \Big]_{ij}(t)\,,\label{eq:EOM_1pFluctuations_PA}
\end{align}
where $\delta\hat{\Sigma}^\mathrm{HF}$ denotes the fluctuations Hartree--Fock selfenergy operator, which follows directly from Eq.~\eqref{eq:def_fluctuations_Hartree_selfenergy} by the replacement $w\rightarrow w^\pm$. \\
This approximation, at the level of the single-particle fluctuations, directly leads to the following set of EOMs for two-time two-particle fluctuations:
\begin{align}
    \mathrm{i}\hbar\partial_t L_{ijkl}(t,t')&= \Big[h^\mathrm{HF},L\Big]^{(1)}_{ijkl}(t,t')+\pi^{(1),\pm}_{ijkl}(t,t')\,,\label{eq:EOM_L_PA_1}\\
     \mathrm{i}\hbar\partial_{t'} L_{ijkl}(t,t')&= \Big[h^\mathrm{HF},L\Big]^{(2)}_{ijkl}(t,t')+\pi^{(2),\pm}_{ijkl}(t,t')\,,\label{eq:EOM_L_PA_2}
\end{align}
where we introduced two-time Hartree--Fock terms of the form
\begin{align}
    \Big[h^\mathrm{HF},L\Big]^{(1)}_{ijkl}(t,t')\coloneqq\, & h^\mathrm{HF}_{ip}(t)L_{pjkl}(t,t')- h^\mathrm{HF}_{pk}(t)L_{ijpl}(t,t')\,, \label{eq:def_HF_L_1}\\
    \Big[h^\mathrm{HF},L\Big]^{(2)}_{ijkl}(t,t')\coloneqq\, & h^\mathrm{HF}_{jp}(t')L_{ipkl}(t,t')- h^\mathrm{HF}_{pl}(t')L_{ijkp}(t,t')\,, \label{eq:def_HF_L_2}
\end{align}
and two-time polarization terms defined as
\begin{align}
    \pi^{(1),\pm}_{ijkl}(t,t')\coloneqq\,&\pm \mathrm{i}\hbar L_{rjpl}(t,t') \big\{ w_{ipqr}^\pm(t)G_{qk}^<(t) -w^\pm_{qpkr}(t)G^<_{iq}(t)\big\}\,, \label{eq:def_pi_1}\\
    \pi^{(2),\pm}_{ijkl}(t,t')\coloneqq\,&\pm \mathrm{i}\hbar L_{iqkp}(t,t') \big\{ w_{pjqr}^\pm(t')G_{rl}^<(t') -w^\pm_{prql}(t')G^<_{jr}(t')\big\}\,. \label{eq:def_pi_2}
\end{align}
Since it holds $\delta\hat{G}_{ij}(t)=-[\delta\hat{G}_{ji}(t)]^\dagger$ and thus $L_{ijkl}(t,t')=[L_{lkji}(t',t)]^*$, the polarization terms obey
\begin{equation}
    \pi^{(1),\pm}_{ijkl}(t,t')= -\Big[\pi^{(2),\pm}_{lkji}(t',t)\Big]^*\,,\label{eq:symmetry}
\end{equation}
and the same symmetry holds for the two-time Hartree--Fock terms, cf. Eqs.~\eqref{eq:def_HF_L_1} and \eqref{eq:def_HF_L_2}.\\
As was shown in Ref.~\cite{schroedter_cmp_22}, the PA is -- in the weak coupling limit -- equivalent to the time-local $GW^\pm$ approximation of the G1--G2 scheme. Equivalence to the $GW$ approximation without said exchange contributions can be established when considering $\delta\hat{\Sigma}^\mathrm{H}$, instead of $\delta\hat{\Sigma}^\mathrm{HF}$, in the EOM for single-particle fluctuations, cf. Eq.~\eqref{eq:EOM_1pFluctuations_PA}. Correspondingly, the EOMs for two-time two-particle fluctuations then include polarization terms $\pi^{(m)}$ which follow from $\pi^{(m),\pm}$ by the replacement $w^\pm\rightarrow w$.

While the polarization approximation is an obvious decoupling ansatz of the fluctuations hierarchy, it remains to understand its physical content and to establish its relations to other approximation schemes. In the following, we establish the close link of the present results to those of the independent Bethe--Salpeter approach of NEGF theory.

\section{Two-Time $GW$ Approximation} \label{s:GWA}
\subsection{Bethe--Salpeter Approach} \label{ss:BSA}
Within the Bethe--Salpeter approach, the XC function obeys the Bethe--Salpeter equation (BSE) on the Keldysh contour
\begin{align}
    L_{ijkl}(z_1,z_2,z'_1,z'_2)=&\pm G_{il}(z_1,z'_2)G_{jk}(z_2,z'_1)+\int_\mathcal{C}G_{ip}(z_1,z_3) G_{rk}(z_5,z'_1) K_{pqrs}(z_3,z_4,z_5,z_6)\times\nonumber \\
    & L_{sjql}(z_6,z_2,z_4,z'_2)\,\mathrm{d}(z_3,\dots,z_6)\,, \label{eq:BSE_L}
\end{align}
where $K$ denotes the two-particle irreducible vertex defined as \cite{kwong_prl_00,stefanucci-book}
\begin{equation}
    K_{ijkl}(z_1,z_2,z'_1,z'_2) \coloneqq \frac{\delta \Sigma_{ik}(z_1,z'_1)}{\delta G_{lj}(z'_2,z_2)}\,, \label{eq:def_irreducible_vertex}
\end{equation}
i.e., as the functional derivative of the single-particle selfenergy $\Sigma$ with respect to the single-particle Green function $G$. Further, we introduce the two-particle reducible vertex $K^\mathrm{red}$, which obeys the following BSE
\begin{align}
    K^\mathrm{red}_{ijkl}(z_1,z_2,z'_1,z'_2) =& \pm K_{ijkl}(z_1,z_2,z'_1,z'_2)+\int_\mathcal{C}K_{ipkq}(z_1,z_3,z'_1,z_4) G_{qr}(z_4,z_5)G_{sp}(z_6,z_3)\times\nonumber\\
    & K^\mathrm{red}_{rjsl}(z_5,z_2,z_6,z'_2)\,\mathrm{d}(z_3,\dots,z_6)\,. \label{eq:BSE_K_red}
\end{align}
Using the reducible vertex, one can express the correlated part of the two-particle NEGF in the following form
\begin{align}
    \mathcal{G}_{ijkl}(z_1,z_2,z'_1,z'_2) = & \int_\mathcal{C} G_{ip}(z_1,z_3) G_{jq}(z_2,z_4)
     K^\mathrm{red}_{pqrs}(z_3,z_4,z_5,z_6) G_{rk}(z_5,z'_1) G_{sp}(z_6,z'_2)\,\mathrm{d}(z_3,\dots,z_6)\,.\label{eq:G2_K_red}
\end{align}
Moreover, the reducible vertex allows us to eliminate any dependence on the XC function on the r.h.s. of Eq.~\eqref{eq:BSE_L}. 
\subsection{$GW$ Approximation within the Bethe--Salpeter Approach}
Within the $GW$ approximation, the irreducible vertex is approximated by 
\begin{align}
    K_{ijkl}(z_1,z_2,z'_1,z'_2)\approx &\pm \mathrm{i}\hbar w_{ijkl}(z_1)\delta_\mathcal{C}(z_1,z_2)\delta_\mathcal{C}(z_1,z'_1)\delta_\mathcal{C}(z_1,z'_2)\,,
\end{align}
i.e., the selfenergy is assumed to be at the level of the Hartree approximation, cf. Eq.~\eqref{eq:def_irreducible_vertex}. Alternatively, one can consider the Hartree--Fock selfenergy for the calculation of the irreducible vertex. This leads to the replacement $w\rightarrow w^\pm$ and gives rise to the $GW^\pm$ approximation, thus restoring exchange contributions that are present in the PA. Consequently, the following considerations can be done analogously for both the $GW$ and the $GW^\pm$ approximation. \\
Within the $GW$ approximation, the BSE for the reducible vertex, cf. Eq.~\eqref{eq:BSE_K_red}, takes the form
\begin{align}
    K^\mathrm{red}_{ijkl}(z_1,z_2,z'_1,z'_2)=\,& \mathrm{i}\hbar w_{ijkl}(z_1) \delta_\mathcal{C}(z_1,z_2)\delta_\mathcal{C}(z_1,z'_1)\delta_\mathcal{C}(z_1,z'_2)\pm\mathrm{i}\hbar\delta_\mathcal{C}(z_1,z'_1) \int_\mathcal{C} w_{ipkq}(z_1) G_{qr}(z_1,z_3) \times\nonumber\\ & G_{sp}(z_4,z_1)
   K^\mathrm{red}_{rjsl}(z_3,z_2,z_4,z'_2) \,\mathrm{d}(z_3,z_4) \,. \label{eq:BSE_K_red_GW}
\end{align}

To establish the link to the present quantum fluctuations approach, we need to reduce the four-time exchange correlation function on the Keldysh contour, cf. Eq.~(\ref{eq:BSE_L}), to a two-time function of real time arguments. We now demonstrate that this can be achieved with help of the Generalized Kadanoff-Baym ansatz (GKBA).

\subsection{Generalized Kadanoff--Baym Ansatz (GKBA)} \label{ss:GKBA}
Within the GKBA \cite{Lipavsky1986}, the time-off-diagonal lesser and greater NEGF are reconstructed from their time-diagonal values using the retarded and advanced Green functions, i.e., 
\begin{equation}
    G^\gtrless_{ij}(t,t') =\mathrm{i}\hbar \Big\{ G^\mathrm{R}_{ik}(t,t')G^\gtrless_{kj}(t')-G^\gtrless_{ik}(t)G^\mathrm{A}_{kj}(t,t')\Big\}\,,
\end{equation}
which is an approximation, see e.g. Ref.~\cite{schluenzen_jpcm_19}.
However, this reconstruction requires knowledge of the retarded and the advanced NEGFs, and one, therefore, needs additional approximations that allow for the calculation of these components without having to solve the general two-time equations. This is achieved within the Hartree--Fock GKBA (HF-GKBA) by treating the real-time components of the single-particle NEGF on the time off-diagonal on the Hartree--Fock level. Effectively, the greater and lesser Green functions then obey simple Hartree--Fock-type equations, i.e., 
\begin{align}
    \mathrm{i}\hbar\partial_t G^\gtrless_{ij}(t,t')&= h^\mathrm{HF}_{ik}(t)G^\gtrless_{kj}(t,t')\,, \label{eq:HF_KBE_1}\\
    -\mathrm{i}\hbar\partial_{t'}G^\gtrless_{ij}(t,t')&=G^\gtrless_{ik}(t,t')h^\mathrm{HF}_{kj}(t')\,,\label{eq:HF_KBE_2}
\end{align}
where Eq.~\eqref{eq:HF_KBE_1} holds for $t\geq t'$ and Eq.~\eqref{eq:HF_KBE_2} for $t\leq t'$. Moreover, the time-diagonal Green functions $G^\gtrless(t)$ are used as initial values for each EOM.
\subsection{Equations of Motion for the Two-Time $GW$ Approximation} \label{ss:EOMs_GW}
In order to derive the EOMs for the two-time $GW$ approximation and show the relation to the two-time PA, we consider the representation of the correlated part of the NEGF in terms of the reducible vertex, cf. Eq.~\eqref{eq:G2_K_red}, in combination with the BSE for the vertex within the $GW$ approximation, cf. Eq.~\eqref{eq:BSE_K_red_GW}. Subsequently, we calculate the time derivatives of the real-time component $\mathcal{G}(t,t')$, cf. Eq.~\eqref{eq:def_2time-G2}, within the HF-GKBA. \\
First of all, we introduce the following notation for integrals of functions $f$ on the contour over the causal and anti-causal branches, $\mathcal{C}_{+/-}$,
\begin{align}
    \int_\alpha f^\alpha(t_\alpha) \coloneqq \alpha\int_{-\infty}^\infty f^\alpha(t)\,\mathrm{d}t\equiv \int_{\mathcal{C}_\alpha} f(z)\,\mathrm{d}z \,, \label{eq:notation_contour_integral}
\end{align}
where $\alpha\in\{+,-\}$. Further, the delta distribution on the Keldysh contour takes the form
\begin{equation}
    \delta_\mathcal{C}(z,z')\equiv \alpha\delta_{\alpha\alpha'} \delta(t,t')\,.
\end{equation}
It follows that two-time two-particle correlations can then be expressed as
\begin{align}
   &\mathcal{G}_{ijkl}(t,t') = \int_{\alpha\beta\gamma\delta} G^{-\alpha}_{ip}(t,t_\alpha)G^{+\beta}_{jq}(t',t_\beta)K^{\mathrm{red},\alpha\beta\gamma\delta}_{pqrs}(t_\alpha,t_\beta,t_\gamma,t_\delta) G^{\gamma-}_{rk}(t_\gamma,t) G^{\delta+}_{sl}(t_\delta,t')\,. \label{eq:2t_G2_K_red}
\end{align}
The Keldysh contour $\mathcal{C}$ does, by construction, not explicitly depend on the times $t$ and $t'$. However, cancellation effects lead to the integrals effectively only being calculated over the intervals $(-\infty,t)$ and $(-\infty,t')$ instead of $(-\infty,\infty)$. Thus, when considering the derivatives with respect to $t$ and $t'$, there are contributions arising due to the integration boundaries and contributions due to the integrands. The latter can be evaluated by means of the HF-GKBA. For $t\geq t'$, we have, within the HF-GKBA, 
\begin{align}
    &\int_{\beta\gamma\delta}\int_{-\infty}^{\Tilde{t}} \big\{ \mathrm{i}\hbar \partial_t G^>_{ip}(t,\Bar{t})\big\}G^{+\beta}_{jq}(t',t_\beta) K^{\mathrm{red},+\beta\gamma\delta}_{pqrs}(\overline{t},t_\beta,t_\gamma,t_\delta) G^{\gamma-}_{rk}(t_\gamma,t) G^{\delta+}_{sl}(t_\delta,t')\,\mathrm{d}\Bar{t} \nonumber \\ \nonumber
    =&\int_{\beta\gamma\delta}\int_{-\infty}^{\Tilde{t}} \big\{ h_{iu}^\mathrm{HF}(t) G^>_{up}(t,\Bar{t})\big\}G^{+\beta}_{jq}(t',t_\beta)K^{\mathrm{red},+\beta\gamma\delta}_{pqrs}(\overline{t},t_\beta,t_\gamma,t_\delta) G^{\gamma-}_{rk}(t_\gamma,t) G^{\delta+}_{sl}(t_\delta,t')\,\mathrm{d}\Bar{t}\,,
\end{align}
where $\Tilde{t}\in\{t,t'\}$ and thus $t\geq \Tilde{t}$. It follows that the contributions due to the derivatives of the integrands of the appearing integrals lead, within the HF-GKBA, to Hartree--Fock contributions similar to Eqs.~\eqref{eq:def_HF_L_1} and \eqref{eq:def_HF_L_2}, i.e., 
\begin{align}
    \Big[h^\mathrm{HF},\mathcal{G}\Big]^{(1)}_{ijkl}(t,t')=\, & h^\mathrm{HF}_{ip}(t)\mathcal{G}_{pjkl}(t,t')- h^\mathrm{HF}_{pk}(t)\mathcal{G}_{ijpl}(t,t')\,, \label{eq:def_HF_G2_1}\\
    \Big[h^\mathrm{HF},\mathcal{G}\Big]^{(2)}_{ijkl}(t,t')=\, & h^\mathrm{HF}_{jp}(t')\mathcal{G}_{ipkl}(t,t')- h^\mathrm{HF}_{pl}(t')\mathcal{G}_{ijkp}(t,t')\,, \label{eq:def_HF_G2_2}
\end{align}
where Eq.~\eqref{eq:def_HF_G2_1} holds for $t\geq t'$ and Eq.~\eqref{eq:def_HF_G2_2} for $t\leq t'$.\\
Next, we consider the terms due to the integral boundaries whose derivation can be split in two parts: the contributions due to the first term on the r.h.s. of Eq.~\eqref{eq:BSE_K_red_GW}, i.e., $\mathrm{i}\hbar w$, and the second term on the r.h.s., i.e., the integral including the reducible vertex. For the first part, we consider the following quantity 
\begin{align}
    \Gamma_{ijkl}(t,t')\coloneqq&\mathrm{i}\hbar \int_\alpha G^{-\alpha}_{ip}(t,t_\alpha) G^{+\alpha}_{jq}(t',t_\alpha) w_{pqrs}(t_\alpha) G^{\alpha-}_{rk}(t_\alpha,t)G^{\alpha+}_{sl}(t_\alpha,t')\,. \label{eq:def_Gamma}
\end{align}
The contribution to $\Gamma$ associated with the causal branch, $\mathcal{C}_+$, can be equivalently expressed as 
\begin{align}
    \int_{-\infty}^\infty G^>_{ip}(t,\Bar{t})G^\mathrm{c}_{jq}(t',\Bar{t}) w_{pqrs}(\Bar{t}) G^<_{rk}(\Bar{t},t)G^\mathrm{c}_{sl}(\Bar{t},t')\,\mathrm{d}\Bar{t}
    =&\int_{-\infty}^{t'} G^>_{ip}(t,\Bar{t})G^>_{jq}(t',\Bar{t}) w_{pqrs}(\Bar{t})G^<_{rk}(\Bar{t},t) G^<_{sl}(\Bar{t},t')\,\mathrm{d}\Bar{t}\nonumber \\&+\int_{t'}^\infty G^>_{ip}(t,\Bar{t}) G^<_{jq}(t',\Bar{t}) w_{pqrs}(\Bar{t}) G^<_{rk}(\Bar{t},t)G^>_{sl}(\Bar{t},t')\,\mathrm{d}\Bar{t} \,. \label{eq:Gamma_+}
\end{align}
Further, the contribution to $\Gamma$ associated with the anti-causal branch, $\mathcal{C}_-$, can be rewritten as
\begin{align}
    \int_{-\infty}^\infty G^\mathrm{a}_{ip}(t,\Bar{t}) G^<_{jq}(t',\Bar{t})w_{pqrs}(\Bar{t}) G^\mathrm{a}_{sl}(\Bar{t},t)G^>_{sl}(\Bar{t},t')\,\mathrm{d}\Bar{t}
    =& \int_{-\infty}^t G^<_{ip}(t,\Bar{t}) G^<_{jq}(t',\Bar{t}) w_{pqrs}(\Bar{t}) G^>_{rk}(\Bar{t},t) G^>_{sl}(\Bar{t},t')\,\mathrm{d}\Bar{t}\nonumber \\
    &+\int_t^\infty G^>_{ip}(t,\Bar{t}) G^<_{jq}(t',\Bar{t}) w_{pqrs}(\Bar{t}) G^<_{rk}(\Bar{t},t) G^>_{sl}(\Bar{t},t')\,\mathrm{d}\Bar{t}\,. \label{eq:Gamma_-}
\end{align}
It follows that $\Gamma$ can be expressed as 
\begin{align}
    \frac{1}{\mathrm{i}\hbar}\Gamma_{ijkl}(t,t') =& \int_{-\infty}^{t'} G^>_{ip}(t,\Bar{t})G^>_{jq}(t',\Bar{t}) w_{pqrs}(\Bar{t})G^<_{rk}(\Bar{t},t) G^<_{sl}(\Bar{t},t')\,\mathrm{d}\Bar{t}\nonumber \\
    &- \int_{-\infty}^t G^<_{ip}(t,\Bar{t}) G^<_{jq}(t',\Bar{t}) w_{pqrs}(\Bar{t}) G^>_{rk}(\Bar{t},t) G^>_{sl}(\Bar{t},t')\,\mathrm{d}\Bar{t}\nonumber \\
    &+\int_{t'}^t G^>_{ip}(t,\Bar{t}) G^<_{jq}(t',\Bar{t}) w_{pqrs}(\Bar{t}) G^<_{rk}(\Bar{t},t) G^>_{sl}(\Bar{t},t')\,\mathrm{d}\Bar{t}\,. \label{eq:Gamma}
\end{align}
Therefore, we find the following contributions due to the integration boundaries (including a prefactor of $\mathrm{i}\hbar$)
\begin{align}
        &\Psi^{(1)}_{ijkl}(t,t')\coloneqq-\hbar^2\big\{G^>_{ip}(t)G^<_{jq}(t',t)w_{pqrs}(t)G^<_{rk}(t)G^>_{sl}(t,t')- G^<_{ip}(t)G^<_{jq}(t',t)w_{pqrs}(t)G^>_{rk}(t)G^>_{sl}(t,t')\big\}\label{eq:def_2t_SOA_1}\,,\\
    &\Psi^{(2)}_{ijkl}(t,t')\coloneqq -\hbar^2 \big\{  G^>_{ip}(t,t') G^>_{jq}(t')w_{pqrs}(t') G^<_{rk}(t',t) G^<_{sl}(t')- G^>_{ip}(t,t') G_{jq}^<(t')w_{pqrs}(t')G^<_{rk}(t',t)G^>_{sl}(t')\big\} \,,\label{eq:def_2t_SOA_2t_2}
\end{align}
where $\Psi^{(1)}$ follows from differentiating Eq.~\eqref{eq:Gamma} with respect to $t$ and $\Psi^{(2)}$ follows from the derivative with respect to $t'$. Using Eq.~\eqref{eq:G</>_property}, we can simplify these expressions and find 
\begin{align}
    \Psi^{(1)}_{ijkl}(t,t')&=  \mathrm{i}\hbar\big\{ w_{ipqr}(t) G^>_{rl}(t,t')G^<_{jp}(t',t) G^<_{qk}(t)-w_{pqkr}(t)G^>_{rl}(t,t')G^<_{jq}(t',t) G^<_{ip}(t)\big\}\,, \label{eq:Psi_1} \\
    \Psi^{(2)}_{ijkl}(t,t')&=  \mathrm{i}\hbar\big\{ w_{pjqr}(t') G^>_{ip}(t,t')G^<_{qk}(t',t) G^<_{rl}(t')-w_{pqrl}(t')G^>_{ip}(t,t')G^<_{rk}(t',t) G^<_{jq}(t')\big\}\,.\label{eq:Psi_2}
\end{align}
For the second part of the derivation, we consider the following quantities
\begin{align}
    \Delta_{ijkl}(t,t') &\coloneqq \int_{\alpha\beta\gamma\delta} G^{-\alpha}_{ip}(t,t_\alpha)G^{+\beta}_{jq}(t',t_\beta)\Phi^{\alpha\beta\gamma\delta}_{pqrs}(t_\alpha,t_\beta,t_\gamma,t_\delta) G^{\gamma-}_{rk}(t_\gamma,t) G^{\delta+}_{sl}(t_\delta,t')\,, \label{eq:def_Delta}\\
    \Phi^{\alpha\beta\gamma\delta}_{ijkl}(t_1,t_2,t_3,t_4) &\coloneqq \pm \mathrm{i}\hbar\alpha \delta_{\alpha\gamma}\delta(t_1,t_3) \int_{\epsilon\zeta} w_{ipkq}(t_1)G^{\alpha\epsilon}_{qr}(t_1,t_\epsilon)K^{\mathrm{red},\epsilon\beta\zeta\delta}_{rjsl}(t_\epsilon,t_2,t_\zeta,t_4) G^{\zeta\alpha}_{sp}(t_\zeta,t_1)\,,\label{eq:def_Phi}
\end{align}
where $\Phi$ denotes the vertex which is reducible in the transversal particle-hole channel. Further, we define 
\begin{align}
        \Omega^{\alpha;\beta \gamma}_{ijkl}(t_1,t_2,t)\coloneqq \int\limits_{\delta\epsilon} G^{\alpha \delta}_{jp}(t,t_\delta) \Phi^{\beta\delta\gamma\epsilon}_{ipkq}(t_1,t_\delta,t_2,t_\epsilon) G^{\epsilon\alpha}_{ql}(t_\epsilon,t)\,. \label{eq:def_Omega}
\end{align}
Utilizing the exchange symmetry of the vertex $\Phi$ of the form $    \Phi_{ijkl}^{\alpha\beta\gamma\delta}(t_1,t_2,t_3,t_4) = \Phi^{\beta\alpha\delta\gamma}_{jilk}(t_2,t_1,t_4,t_3)$,
we find a more compact form of $\Delta$ given by
\begin{align}
    \Delta_{ijkl}(t,t')&= \int_{\alpha\gamma} G^{-\alpha}_{ip}(t,t_\alpha) \Omega^{+;\alpha\gamma}_{pjrl}(t_\alpha,t_\gamma,t') G^{\gamma -}_{rk}(t_\gamma,t) \label{eq:Omega_+} \\
        &=\int_{\beta\delta} G^{+\beta}_{jq}(t',t_\beta) \Omega^{-;\beta\delta}_{qisk}(t_\beta,t_\delta,t) G^{\delta +}_{sl}(t_\delta,t')\,. \label{eq:Omega_-}
\end{align}
Let us first consider the derivative of $\Delta$ with respect to $t$. Here, we only have to consider the case $\alpha=\gamma$ in Eq.~\eqref{eq:Omega_+} due to the delta distribution appearing  in Eq.~\eqref{eq:def_Phi}. For $\alpha=\gamma=-$, we find
\begin{align}
 \int_{-\infty}^\infty G^\mathrm{a}_{ip}(t,\Bar{t}) \Omega^{+;--}_{pjrl}(\Bar{t},\Bar{t},t') G^\mathrm{a}_{rk}(\Bar{t},t)\,\mathrm{d}\Bar{t}
    =&\int_{-\infty}^t G^<_{ip}(t,\Bar{t}) \Omega^{+;--}_{pjrl}(\Bar{t},\Bar{t},t')G^>_{rk}(\Bar{t},t)\,\mathrm{d}\Bar{t}\nonumber \\&+ \int_t^\infty G^>_{ip}(t,\Bar{t}) \Omega^{+;--}_{pjrl}(\Bar{t},\Bar{t},t') G^<_{rk}(\Bar{t},t)\,\mathrm{d}\Bar{t}\,.
\end{align}
Using the symmetry $\Phi^{+\beta+\delta}_{ijkl}(t_1,t_2,t_3,t_4)=-\Phi^{-\beta-\delta}_{ijkl}(t_1,t_2,t_3,t_4)$ \cite{joost_phd_2022}, the second integral on the r.h.s. over the interval $(t,\infty)$ is canceled by the contributions with $\alpha =\gamma =+$. Hence, we have 
\begin{align}
    \Delta_{ijkl}(t,t') =& \int_{-\infty}^t G^<_{ip}(t,\Bar{t}) \Omega^{+;--}_{pjrl}(\Bar{t},\Bar{t},t')G^>_{rk}(\Bar{t},t)\,\mathrm{d}\Bar{t}- \int_{-\infty}^t G^>_{ip}(t,\Bar{t}) \Omega^{+;--}_{pjrl}(\Bar{t},\Bar{t},t') G^<_{rk}(\Bar{t},t)\,\mathrm{d}\Bar{t}\,.
\end{align}
The integration boundaries thus give rise to the term\footnote{Here, we again include a pre-factor of $\mathrm{i}\hbar$ and use the property (\ref{eq:G</>_property}).}
\begin{align}
    \Pi^{(1)}_{ijkl}(t,t') \coloneqq \big\{ G^<_{ip}(t)\delta_{rk}-\delta_{ip} G^<_{rk}(t) \big\} \Omega_{pjrl}^{+;--}(t,t,t') \,. \label{eq:def_Pi_G2_1.1}
\end{align}
Lastly, it remains to further investigate $\Omega^{+;--}_{pjrl}(t,t,t')$ in order to identify $\mathcal{G}$. Here, it follows 
\begin{align}
    \Omega_{pjrl}^{+;--}(t,t,t')&= \mp \mathrm{i}\hbar \int_{\beta\delta\epsilon\zeta} G^{+\beta}_{jq}(t',t_\beta) w_{purv}(t)  G^{-\epsilon}_{vx}(t,t_\epsilon) K^{\mathrm{red},\epsilon\beta\zeta\delta}_{xqys}(t_\epsilon,t_\beta,t_\zeta,t_\delta) G^{\zeta-}_{yu}(t_\zeta,t) G^{\delta+}_{sl}(t_\delta, t')\\
    &= \mp \mathrm{i}\hbar w_{purv}(t) \mathcal{G}_{vjul}(t,t') \,.
\end{align}
Consequently, we find the following expression for $\Pi^{(1)}$: 
\begin{align}
    \Pi^{(1)}_{ijkl}(t,t')\coloneqq\,&\pm \mathrm{i}\hbar\mathcal{G}_{rjpl}(t,t') \big\{ w_{ipqr}(t)G_{qk}^<(t) -w_{qpkr}(t)G^<_{iq}(t)\big\}\,, \label{eq:def_Pi_G2_1.2}
\end{align}
which has the same structure as $\pi^{(1)}$ when replacing $L\leftrightarrow\mathcal{G}$, cf. Eq.~\eqref{eq:def_pi_1} (with $w^\pm\rightarrow w$). In fact, using Eq.~\eqref{eq:L_G2}, we find that 
\begin{equation}
    \pi^{(1)}_{ijkl}(t,t')= \Psi^{(1)}_{ijkl}(t,t')+\Pi^{(1)}_{ijkl}(t,t')\,. 
\nonumber
\end{equation}
Now, utilizing the exchange symmetries of $\Phi$ and the appearing deltas, we only have to consider the cases with $\beta=\delta$ in Eq.~\eqref{eq:Omega_-}. Analogously to the previous derivation, we find that the boundary terms lead to contributions of the form 
\begin{align}
     \Pi^{(2)}_{ijkl}(t,t')\coloneqq\,&\pm \mathrm{i}\hbar \mathcal{G}_{iqkp}(t,t') \big\{ w_{pjqr}(t')G_{rl}^<(t') -w_{prql}(t')G^<_{jr}(t')\big\}\,, \label{eq:def_Pi_G2_2}
\end{align}
where we again recognize the same structure as for $\pi^{(2)}$ and find 
\begin{equation}
    \pi^{(2)}_{ijkl}(t,t')=\Psi^{(2)}_{ijkl}(t,t')+\Pi^{(2)}_{ijkl}(t,t')\,. 
\nonumber
\end{equation}
In summary, we find for the two-time $GW$ approximation the following set of EOMs
\begin{align}
    \mathrm{i}\hbar \partial_t \mathcal{G}_{ijkl}(t,t')&=\Big[h^\mathrm{HF},\mathcal{G}\Big]^{(1)}_{ijkl}(t,t')+\Psi^{(1)}_{ijkl}(t,t')+\Pi^{(1)}_{ijkl}(t,t')\,,\label{eq:EOM_G2_GW_1}\\
     \mathrm{i}\hbar\partial_{t'} \mathcal{G}_{ijkl}(t,t')&= \Big[h^\mathrm{HF},\mathcal{G}\Big]^{(2)}_{ijkl}(t,t')+\Psi^{(2)}_{ijkl}(t,t')+\Pi^{(2)}_{ijkl}(t,t')\,.\label{eq:EOM_G2_GW_2}
\end{align}
Exchange contributions similar to those included in the PA follow from the replacement $w\rightarrow w^\pm$, in $\Psi^{(m)}$ and $\Pi^{(m)}$, giving rise to $\Psi^{(m),\pm}$ and $\Pi^{(m),\pm}$.
\section{Equivalence of the PA and the $GW^\pm$ Approximation} \label{s:equivalence}
As we have seen in the previous section, the EOMs for $L$, in the PA, on the one hand, and for $\mathcal{G}$, in the $GW^{(\pm)}$ approximation, on the other hand, display a close resemblance. However, the former is expressed in terms of fluctuations while the latter is expressed in terms of correlations. Thus, in order to show the equivalence of the two approximations, it is necessary to consider the EOMs for source fluctuations, $L^{(0)}$, as it holds $L=L^{(0)}+\mathcal{G}$, cf. Eq.~\eqref{eq:L_G2}. Recall that source fluctuations are defined as a product of single-particle Green functions, i.e., $L^{(0)}(t,t')=\pm G^>(t,t')G^<(t',t)$, cf. Eq.~\eqref{eq:def_source_fluctuations}. Hence, the EOMs for $L^{(0)}$ solely depend on the EOMs for $G^\gtrless$ on the time off-diagonal. Consequently, within the HF-GKBA, cf. Eqs.~\eqref{eq:HF_KBE_1} and \eqref{eq:HF_KBE_2}, the EOMs for source fluctuations are given by
\begin{align}
    \mathrm{i}\hbar\partial_tL^{(0)}_{ijkl}(t,t')&=\Big[h^\mathrm{HF},L^{(0)}\Big]^{(1)}_{ijkl}(t,t')\,, \label{eq:L0_1_EOM}\\
    \mathrm{i}\hbar\partial_{t'}L^{(0)}_{ijkl}(t,t')&=\Big[h^\mathrm{HF},L^{(0)}\Big]^{(2)}_{ijkl}(t,t')\,. \label{eq:L0_2_EOM}
\end{align}
This directly implies that the two-time EOMs in the PA are the same as those in the $GW^\pm$ approximation.\\
Further, it is necessary to consider the boundary conditions of the EOMs in the respective approximations to show their equivalence. These are determined by the time-diagonal functions $L(t)$ or $\mathcal{G}(t)$. Hence, the EOMs on the time diagonal have to be considered. In the $GW^\pm$ approximation, the EOM for two-particle correlations, $\mathcal{G}^{GW^\pm}$, is given by 
\begin{equation}
    \mathrm{i}\hbar \partial_t \mathcal{G}^{GW^\pm}_{ijkl}(t) = \Big[h^{(2),\mathrm{HF}}, \mathcal{G}^{GW^\pm} \Big]_{ijkl}(t) + \Psi^\pm_{ijkl}(t)+\Pi^\pm_{ijkl}(t)\,, \label{eq:EOM_G2_GW_TD}
\end{equation}
where the terms on the r.h.s. follow from the two-time quantities defined in Sec.~\ref{s:GWA} and are given by\footnote{The expression given in Eq.~\eqref{eq:two-particle_commutator} corresponds to a commutator of two rank-four tensors defined as $[A,B]_{ijkl}\coloneqq A_{ijpq}B_{pqkl}-B_{ijpq}A_{pqkl}$. Here, we have $h^{(2),\mathrm{HF}}_{ijkl}(t)\coloneqq h^\mathrm{HF}_{ik}(t)\delta_{jl}+h^\mathrm{HF}_{jl}(t)\delta_{ik}$.}
\begin{align}
    \Big[h^{(2),\mathrm{HF}}, \mathcal{G}^{GW^\pm} \Big]_{ijkl}(t)&\coloneqq \Big[h^{\mathrm{HF}}, \mathcal{G}^{GW^\pm} \Big]^{(1)}_{ijkl}(t,t)+\Big[h^{\mathrm{HF}}, \mathcal{G}^{GW^\pm} \Big]^{(2)}_{ijkl}(t,t)\,, \label{eq:two-particle_commutator}\\ 
    \Psi^\pm_{ijkl}(t)&\coloneqq \Psi^{(1),\pm}_{ijkl}(t,t)+\Psi^{(2),\pm}_{ijkl}(t,t)\,,\\
    \Pi^\pm_{ijkl}(t)&\coloneqq \Pi^{(1),\pm}_{ijkl}(t,t)+\Pi^{(2),\pm}_{ijkl}(t,t)\,.
\end{align}
On the other hand, on the time diagonal, two-particle fluctuations in the PA, $L^\mathrm{P}$, obey the following EOM
\begin{equation}
    \mathrm{i}\hbar \partial_t L^\mathrm{P}_{ijkl}(t) = \Big[h^{(2),\mathrm{HF}},L^\mathrm{P}\Big]_{ijkl}(t)+\pi^\pm_{ijkl}(t)\,, \label{eq:EOM_L_P_TD}
\end{equation}
where the polarization term on the r.h.s. is defined as
\begin{equation}
    \pi^\pm_{ijkl}(t)\coloneqq \pi^{(1),\pm}_{ijkl}(t,t)+\pi^{(2),\pm}_{ijkl}(t,t)\,.
\end{equation}
Similar to the time-off-diagonal case, it is now necessary to consider the EOM for source fluctuations to establish the relation between the PA and $GW^\pm$. Here, it follows by utilizing the EOM for $G^<(t)$, cf. Eq.~\eqref{eq:time-diagonal_KBE}, 
\begin{equation}
    \mathrm{i}\hbar \partial_t L^{(0)}_{ijkl}(t)= \Big[h^{(2),\mathrm{HF}},L^{(0)}\Big]_{ijkl}(t)+\mathcal{R}_{ijkl}(t) \,, \label{eq:EOM_L0_TD}
\end{equation}
where the last term on the r.h.s. denotes a residual term that is defined as
\begin{equation}
    \mathcal{R}_{ijkl}(t)\coloneqq \pm G^>_{il}(t)\Big[I+I^\dagger\Big]_{jk}(t)\pm \Big[I+I^\dagger\Big]_{il}(t)G^<_{jk}(t)\,,\label{eq:definition_residual_term}
\end{equation}
which stems from the collision term that is present in the time-diagonal case. In the weak coupling limit, however, (source) fluctuations are significantly larger than correlations, i.e., $|\mathcal{G}|\ll |L|,|L^{(0)}|$, thus making the residual term negligible, i.e., $\mathcal{R}\approx 0$. Hence, the time-diagonal EOM for source fluctuations in the PA under this additional assumption is given by
\begin{equation}
    \mathrm{i}\hbar\partial_t L^{(0),\mathrm{P}}_{ijkl}(t)= \Big[h^{(2),\mathrm{HF}},L^{(0),\mathrm{P}}\Big]_{ijkl}(t)\,. \label{eq:EOM_L0_P_TD}
\end{equation}
It follows that the EOM for two-particle correlations in the PA, $\mathcal{G}^\mathrm{P}= L^\mathrm{P}-L^{(0),\mathrm{P}}$, is given by Eq.~\eqref{eq:EOM_G2_GW_TD}. This implies that the PA and $GW^\pm$ are equivalent on the time diagonal and thus also on the time off-diagonal as the two-time EOMs are the same and have the same boundary conditions.\\

In general, however, the PA, within the fluctuations framework, is used without explicitly considering the EOM for source fluctuations. Consequently, the residual term $\mathcal{R}$ is not explicitly neglected, i.e., two-particle correlations are given by $\mathcal{G}^\mathrm{P}= L^\mathrm{P}-L^{(0)}$ and obey an EOM of the form
\begin{equation}
    \mathrm{i}\hbar \partial_t \mathcal{G}^\mathrm{P}_{ijkl}(t)=\Big[h^{(2),\mathrm{HF}},\mathcal{G}^\mathrm{P}\Big]_{ijkl}(t)+\Psi_{ijkl}^\pm+\Pi^\pm_{ijkl}-\mathcal{R}_{ijkl}(t)\,.
\end{equation}
Thus, in this formulation of the PA, a weaker form of the equivalence of the two approximations holds. As both, however, are weak coupling approximations, the PA and $GW^\pm$ can be considered equivalent within their range of applicability. See Ref.~\cite{schroedter_cmp_22} for more details about the equivalence of the two approximations on the time diagonal.
\section{Application to the Fermi--Hubbard Model} \label{s:application} 
\subsection{The Fermi--Hubbard Hamiltonian}
For the Fermi--Hubbard model, the single-particle contributions to the Hamiltonian, cf. Eq.~\eqref{eq:Hamiltonian}, are described by a hopping term of the form
\begin{equation}
    h^\mathrm{FH}_{ij}\coloneqq -\delta_{\langle i,j\rangle} J\,,
\end{equation}
where $J$ denotes the hopping amplitude between neighboring sites and $\delta_{\langle i,j\rangle}=1$, if the sites $i$ and $j$ are adjacent, and  $\delta_{\langle i,j\rangle}=0$, if they are not. Further, the pair-interaction of the Hamiltonian is given by 
\begin{equation}
    w^{\mathrm{FH},\sigma_1\sigma_2\sigma'_1\sigma'_2}_{ijkl}\coloneqq U \delta_{ij}\delta_{ik}\delta_{il} \delta_{\sigma_1\sigma'_2}\delta_{\sigma_2\sigma'_1}\big\{1-\delta_{\sigma_1\sigma_2}\big\}\,,\label{eq:Hubbard_interaction_matrix}
\end{equation} 
with on-site interaction strength $U$ and the spin components denoted by $\sigma_m,\sigma'_m\in\{\uparrow,\downarrow\}$. This gives rise to the Fermi--Hubbard Hamiltonian of the form
\begin{align}
    \hat{H}_\mathrm{FH}\coloneqq -\delta_{\langle i,j\rangle}J\hat{c}^\dagger_{i\sigma}\hat{c}_{j\sigma}+U \hat{n}_i^\uparrow\hat{n}^\downarrow_i\,, \label{eq:FH-Hamiltonian}
\end{align}
where $\hat{n}^\sigma_i\coloneqq \hat{c}^\dagger_{i\sigma}\hat{c}_{i\sigma}$ describes the density operator on site $i$ with respect to the spin component $\sigma$. It has to be highlighted here that the inclusion of the factor $1-\delta_{\sigma_1\sigma_2}$ on the r.h.s. of Eq.~\eqref{eq:Hubbard_interaction_matrix} is sometimes omitted. Both choices lead to the same Hamiltonian, cf. Eq.~\eqref{eq:FH-Hamiltonian}, and are thus equivalent for the exact description of the system. However, the explicit inclusion of Pauli blocking described by this additional factor can lead to differences when considering approximations. More specifically, this leads to $w^\mathrm{FH}=w^{\mathrm{FH}\pm}$, which is not the case if the blocking factor is not explicitly included.
\subsection{Implementation of the Two-Time PA and $GW$ Approximation} \label{ss:implementation}
As was mentioned before, due to the specific choice of the pair-interaction, we have $w^\mathrm{FH}=w^{\mathrm{FH}\pm}$, and thus, $GW$ and $GW^\pm$ are equivalent. Consequently, we will not differentiate between the two approximations in the following and simply refer to them as $GW$. Furthermore, we assume spin symmetry, i.e.,
\begin{gather}
    G^<_{ij}(t)\equiv G^{<,\uparrow\uparrow}_{ij}(t)\equiv G^{<,\downarrow\downarrow}_{ij}(t)\,,\\
    L^{\sigma_1\sigma_2}_{ijkl}(t,t')\equiv L^{\sigma_1\sigma_2\sigma'_1\sigma'_2}_{ijkl}(t,t')\equiv L^{\overline{\sigma_1\sigma_2\sigma'_1\sigma'_2}}_{ijkl}(t,t') \equiv  L^{\overline{\sigma_1\sigma_2}}_{ijkl}(t,t')\,,\\
    \mathcal{G}^{\sigma_1\sigma_2}_{ijkl}(t,t')\equiv \mathcal{G}^{\sigma_1\sigma_2\sigma'_1\sigma'_2}_{ijkl}(t,t')\equiv \mathcal{G}^{\overline{\sigma_1\sigma_2\sigma'_1\sigma'_2}}_{ijkl}(t,t') \equiv  \mathcal{G}^{\overline{\sigma_1\sigma_2}}_{ijkl}(t,t')\,,
\end{gather}
where $\sigma_m = \uparrow (\downarrow)$ implies $\overline{\sigma_m}=\downarrow (\uparrow)$.\\ The solution of the two-time equations for the PA and the $GW$ approximation requires the solution of the respective sets of EOMs on the time diagonal. Within the Hubbard model, the EOM for the single-particle Green function is given by
\begin{align}
    \mathrm{i}\hbar \partial_t G^<_{ij}(t) = \Big[h^{(1)},G^<\Big]_{ij}(t)+ \Big[ I+I^\dagger \Big]_{ij}(t)\,, \label{eq:EOM_G1_Hubbard}
\end{align}
where $h^{(1)}$ denotes the Hartree--(Fock) Hamiltonian and is given by 
\begin{equation}
    h_{ij}^{(1)}(t)\coloneqq h^\mathrm{FH}_{ij}-\mathrm{i}\hbar \delta_{ij} U G^<_{ii}(t)\,.
\end{equation}
 Moreover, since the single-particle Green function vanishes for differing spin combinations, i.e., $G^{\uparrow\downarrow}=G^{\downarrow\uparrow}=0$, there are no source fluctuations of the form $L^{(0),\uparrow\downarrow}$ and $L^{(0),\downarrow\uparrow}$, and thus, two-particle fluctuations and correlations are equal in this case, i.e., $L^{\uparrow\downarrow}=\mathcal{G}^{\uparrow\downarrow}$. Consequently, the collision term is the same when expressed in terms of fluctuations as well as correlations within the Hubbard model and is given by 
\begin{equation}
    I_{ij}(t)= -\frac{\mathrm{i}\hbar}{2}U\big\{ L^{\uparrow\downarrow}_{iiij}(t)+L^{\uparrow\downarrow}_{iiji}(t) \big\}\,,
\end{equation}
where a symmetrized version of the expression given in Eq.~\eqref{eq:fluc_collision_term} is used. This is attributed to the breaking of exchange symmetry within the PA, causing instabilities in the numerical solution of the equations \cite{schroedter_23}. However, this issue can be resolved by considering a symmetric collision term that inherently possesses this symmetry. Within the $GW$ approximation, this is, in principle, not necessary, and both expressions lead to the same numerical solution. \\
The EOM for two-particle fluctuations on the time diagonal within the PA, $L^\mathrm{P}(t)$, that is being numerically solved, is given by 
\begin{equation}
    \mathrm{i}\hbar \partial_t L^{\mathrm{P},\sigma_1\sigma_2}_{ijkl}(t)= \Big[ h^{(2)}, L^{\mathrm{P},\sigma_1\sigma_2} \Big]_{ijkl}(t)+\pi^{\sigma_1\sigma_2}_{ijkl}(t)\,,
\end{equation}
where we set for the two-particle Hartree--(Fock) Hamiltonian
\begin{equation}
    h^{(2)}_{ijkl}(t)\coloneqq h^{(1)}_{ik}(t)\delta_{jl}+h^{(1)}_{jl}(t)\delta_{ik}\,,
\end{equation}
and have for the polarization contribution in the Hubbard basis
\begin{equation}
    \pi^{\sigma_1\sigma_2}_{ijkl}(t)= -\mathrm{i}\hbar U\Big[ G^<_{ik}(t)\big\{L_{ijil}^{\sigma_1\overline{\sigma_2}}(t)-L_{kjkl}^{\sigma_1\overline{\sigma_2}}(t)\big\}+G^<_{jl}(t)\big\{L^{\sigma_1\overline{\sigma_2}}_{ilkl}(t)-L^{\sigma_1\overline{\sigma_2}}_{ijkj}(t)\big\} \Big]\,.\label{eq:polarization_Hubbard}
\end{equation}
For the $GW$ approximation, we have the following EOMs for the different spin components of two-particle correlations given by 
\begin{align}
    \mathrm{i}\hbar \partial_t \mathcal{G}^{GW,\uparrow\downarrow}_{ijkl}(t)&= \Big[h^{(2)},\mathcal{G}^{GW,\uparrow\downarrow}\Big]_{ijkl}(t)+\Psi^{\uparrow\downarrow}_{ijkl}(t)+\Pi^{\uparrow\downarrow}_{ijkl}(t)\,,\\
    \mathrm{i}\hbar \partial_t \mathcal{G}^{GW,\uparrow\uparrow}_{ijkl}(t)&= \Big[h^{(2)},\mathcal{G}^{GW,\uparrow\uparrow}\Big]_{ijkl}(t)+\Pi^{\uparrow\uparrow}_{ijkl}(t)\,,
\end{align}
where the $\uparrow\uparrow$-component of the second-order contributions vanishes, i.e., $\Psi^{\uparrow\uparrow}=0$, and the terms $\Pi^{\uparrow\downarrow}$ and $\Pi^{\uparrow\uparrow}$ follow from Eq.~\eqref{eq:polarization_Hubbard} by the replacement $L\rightarrow \mathcal{G}$. The non-vanishing second-order contribution is given by
\begin{align}
    \Psi^{\uparrow\downarrow}_{ijkl}(t)= -\mathrm{i}\hbar U\Big[ G^<_{ik}(t)\big\{  G^>_{il}(t)G^<_{ji}(t)-G^>_{kl}(t)G^<_{jk}(t)\big\}+G^<_{jl}(t)\big\{G^>_{il}(t)G^<_{lk}(t)-G^>_{ij}(t)G^<_{jk}(t)\big\} \Big]\,.
\end{align}
Further, the initial values of the EOMs for both approximations are given by $ G^<_{ij}(t_0)= G^{<,0}_{ij}$ and $\mathcal{G}^{\sigma_1\sigma_2}_{ijkl}(t_0)= \mathcal{G}^{\sigma_1\sigma_2,0}_{ijkl}$.\\
We now turn to the two-time equations. Here, both approximations lead to the same EOMs and since, additionally, the symmetries of the different contributions, cf. Eq.~\eqref{eq:symmetry}, can be used, it is only necessary to consider the following EOM on the time off-diagonal for $t\geq t'$,
\begin{equation}
    \mathrm{i}\hbar\partial_t L^{\sigma_1\sigma_2}_{ijkl}(t,t') = \Big[h^{(1)},L^{\sigma_1\sigma_2}\Big]^{(1)}_{ijkl}(t,t')+\pi^{(1),\sigma_1\sigma_2}_{ijkl}(t,t')\,,
\end{equation}
where the two-time polarization term in the Hubbard basis is given by
\begin{equation}
    \pi^{(1),\sigma_1\sigma_2}_{ijkl}(t,t') = - \mathrm{i}\hbar UG^<_{ik}(t) \big\{L^{\sigma_1\overline{\sigma_2}}_{ijil}(t,t')- L^{\sigma_1\overline{\sigma_2}}_{kjkl}(t,t')\big\}\,. 
\end{equation}
For any $t=t'$, the time-diagonal value of the two-particle fluctuations, $L^{\sigma_1\sigma_2}_{ijkl}(t')$, is then used as the initial value of the time-off-diagonal propagation of the two-time fluctuations.\\
To test the two-time approximations, we need to compute an observable that directly depends on two-time two-particle fluctuations or correlations. To this end, we consider the retarded component of the density response function $\chi$ \footnote{The general density response function is defined in terms of the XC function as $\chi_{ij}(z,z')\coloneqq \mathrm{i}\hbar L_{ijij}(z,z',z^+,z^{'+})$.}, which is defined as
\begin{equation}
    \chi^\mathrm{R}_{ij}(t,t') \coloneqq \frac{1}{\mathrm{i}\hbar}\Theta(t-t') \Big\langle\Big[\hat{n}^{\sigma_1}_i(t),\hat{n}^{\sigma_2}_j(t')\Big]\Big\rangle = -4\hbar \mathrm{Im}\Big[ L^{\uparrow\downarrow}_{ijij}(t,t')+L^{\uparrow\uparrow}_{ijij}(t,t') \Big]\,,
\end{equation} 
where the exchange symmetries of the two-particle fluctuations as well as the spin symmetry were used for the second equality.
\subsection{Numerical Results} \label{ss:numerics}
\begin{figure}
    \centering
    \includegraphics{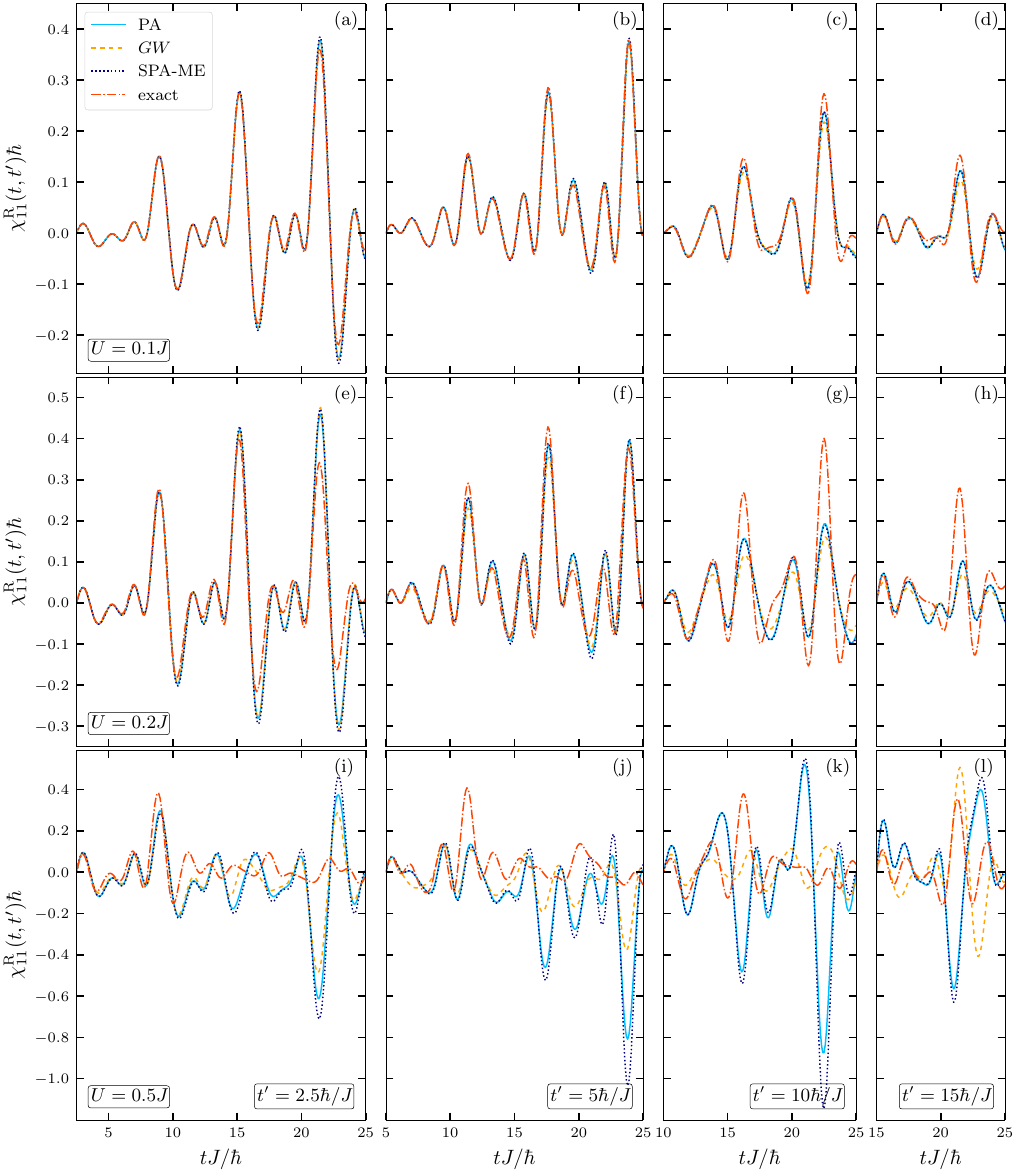}
    \caption{Retarded component of the density response function for a half-filled six-site Hubbard chain with PBC at $U/J=0.1, 0.2, 0.5$ (rows), following a confinement quench at $t=0$, for fixed $t'J/\hbar = 2.5, 5, 10, 15$ (columns). The results for the PA (solid lines) and $GW$ approximation (dashed lines) are compared to the SPA-ME (dotted lines) and exact calculations (dash-dotted lines). }
    \label{fig:chi}
\end{figure}
We now present numerical results, comparing the two-time approximations as well as extending the results for nonequilibrium response functions of Ref.~\cite{schroedter_23} by additionally considering data obtained from exact diagonalization and the stochastic polarization approximation within the multiple ensembles approach (SPA-ME). The latter represents an approach that combines the polarization approximation with stochastic methods to allow for the calculation of spectral two-particle observables, such as the density response function and the dynamic structure factor. More details on this approach and its implementation for the Hubbard model are given in Ref.~\cite{schroedter_23}.\\
Here, we consider half-filled chains with six as well as thirty sites and periodic boundary conditions (PBC) following a confinement quench. Specifically, for a chain with $N$ sites, the sites $1,\dots,N/2$ are fully occupied, while the remaining sites $N/2+1,\dots, N$ are empty. 
At time $t=0$, the confinement is lifted instantaneously, initiating a diffusion-type process in the system, e.g.~\cite{schluenzen_prb16}. The initial state is taken to be uncorrelated, i.e., $\mathcal{G}^{0}=0$. \\

First, we consider the six-site chain and compare the PA and the $GW$ approximation to the SPA-ME and exact diagonalization. Figure~\ref{fig:chi} illustrates the evolution of the retarded component of the density response function, $\chi^\mathrm{R}_{11}(t,t')$, with respect to time $t$ for different interaction strengths $U$ and fixed values of $t'$. In panels (a)-(d), we observe very good agreement among the three approximations for $U=0.1J$, with only minor deviations from the exact solution, which become more pronounced with increasing time ($t$ and $t'$). For $U=0.2J$, in panels (e)-(h), more noticeable differences emerge between the approximations and the exact solution, particularly in the damping behavior observed in the latter that is missing for the approximations. Similar behavior is typical for simulations using HF propagators in the GKBA as those are known 
to underestimate the damping of the dynamics. Although this is not directly visible for $t'J/\hbar=5,10,15$ in panels (f)-(h), due to the short propagation time, it is also expected to occur there for longer times. Notably, SPA-ME and PA show good agreement regarding the frequencies of the oscillations at $U=0.5J$, but deviations in amplitudes become more significant for longer times and larger amplitudes. This behavior is attributed to the semi-classical nature of SPA-ME, which replaces quantum mechanical expectation values with classical stochastic ones, thereby failing to accurately capture certain quantum effects, such as coherence, especially for stronger coupling. Moreover, for larger on-site interaction strength, the amplitudes of the oscillations significantly increase, in the case of the approximations. This trend is observable to a far lesser extent in the exact solution. For $U=0.5J$, the approximations severely overestimate fluctuations, as compared to the exact result. In panel (i), the largest amplitude is in between $\chi^\mathrm{R}\approx-0.5/\hbar$ and $\chi^\mathrm{R}\approx-0.7/\hbar$ at $t=20\hbar/J$ for the approximations, while, concurrently, the exact solution remains bounded within the interval from $-0.2/\hbar$ to $0.2/\hbar$. This discrepancy is consistently observed in panels (j) and (k), where oscillations of the exact solution mostly remain bounded within the same interval. The polarization approximations, however, display peaks of size $\chi^\mathrm{R}\approx -0.8/\hbar$ (PA) and $\chi^\mathrm{R}\approx -1.0/\hbar$ (SPA-ME) in panel (j), as well as $\chi^\mathrm{R}\approx -0.95 /\hbar$ (PA) and $\chi^\mathrm{R}\approx -1.1 /\hbar$ (SPA-ME) in panel (k). In contrast, the $GW$ approximation does not exhibit this trend to the same degree.  \\

Next, for $t'=2.5\hbar/J$ in panel (i), all approximations exhibit excellent agreement up to $t\approx 15\hbar/J$, beyond which differences become more apparent. As $t'$ increases, the range of agreement shifts to smaller time differences $t-t'$. For instance, for $t'=5\hbar/J$ in panel (j), deviations become visible after $t-t'\approx 10 \hbar/J$, and for $t'=10\hbar/J$ in panel (k), after $t-t'\approx 2.5\hbar/J$. However, for $t'=15\hbar/J$ in panel (l), pronounced differences are already visible for $t\approx t'$. Overall, the range of agreement between the approximations decreases exponentially with increasing $t'$. While similar frequencies of the oscillations are observed for $t'=2.5\hbar/J$ and $t'=5\hbar/J$ at larger times $t$, a significant phase shift is noticeable in the oscillations for the $GW$ approximation compared to PA and SPA-ME at $t'=10\hbar/J$ and $t'=15\hbar/J$. This highlights the sensitivity of the system's dynamics to the chosen approximation and the impact of weak perturbations in the initial conditions, which significantly affect the propagation for later times.\\
\begin{figure}
    \centering
    \includegraphics[width=0.95\textwidth]{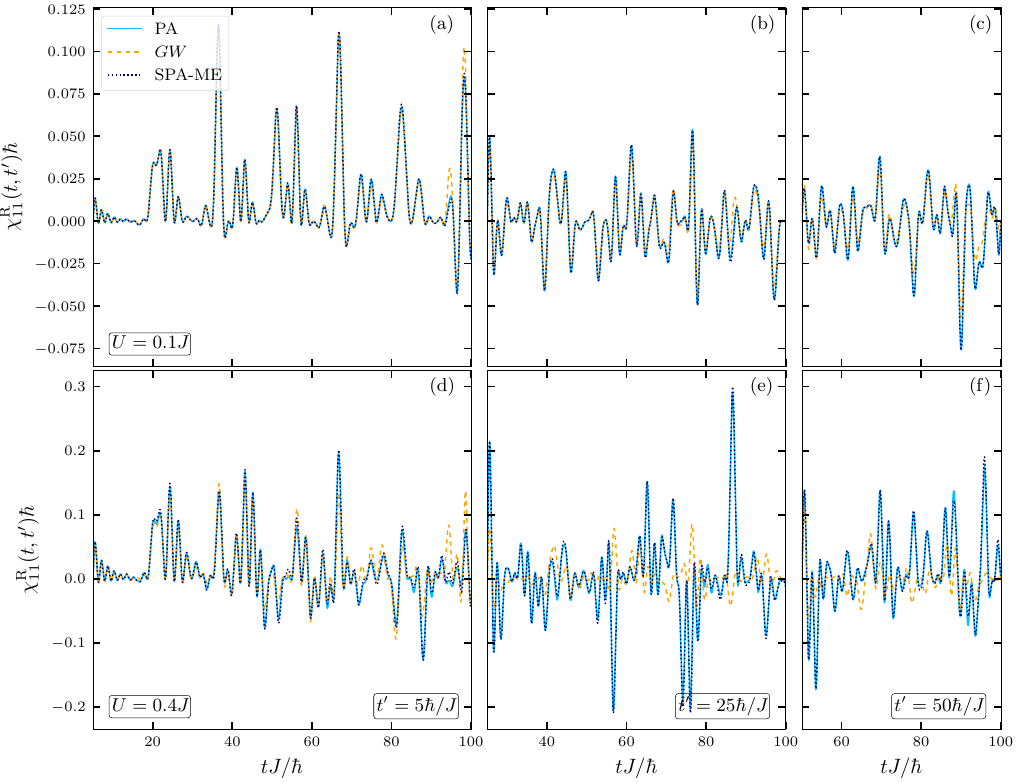}
    \caption{Retarded component of the density response function for a half-filled thirty-site Hubbard chain with PBC at $U/J=0.1, 0.4$ (rows), following a confinement quench at $t=0$, for fixed $t'J/\hbar = 5, 25, 50$ (columns). The results for the PA (solid lines) and $GW$ approximation (dashed lines) are compared to the SPA-ME (dotted lines). }
    \label{fig:chi_30}
\end{figure}
Next, we consider a larger system with thirty sites. Here we compare the two-time approximations, i.e. $GW$ and PA, to the SPA-ME. Figure~\ref{fig:chi_30} shows that the two-time PA and the SPA-ME perfectly agree with each other for both coupling strengths at all times $t$ and $t'$. This illustrates that the effects of introducing the multiple ensembles approach to the PA become negligible for larger systems. For a weakly coupled system at $U=0.1J$, shown in panels (a)-(c), we see excellent agreement between the polarization approximations and the $GW$ approximation for almost all times. In panel (a), deviations between the approximations are only visible for $t\gtrsim 95\hbar/J$ and are limited to the amplitudes of the oscillations for the $GW$ approximation being greater than for the polarization approximations. For later times $t'$, shown in panels (b) and (c), we see minor deviations arising earlier. Nonetheless, agreement between the approximations is very good for almost all times $t$ and $t'$.\\
At $U=0.4J$, shown in panels (d)-(f), similar to the six-site system, larger deviations between the approximations are observable. Panel (d) shows very good agreement of all approximations for times $t\lesssim 70 \hbar/J$ for fixed $t'= 5\hbar/J$. For later times $t$, both the amplitudes as well as the frequencies of the oscillations show increasingly larger deviations. Panels (e) and (f) show that there is mostly qualitative agreement between the approximations even for small times $t$. For $t'=25\hbar/J$, shown in panel (e), we see similar behavior of the oscillations for times $t\lesssim 55\hbar/J$, i.e., relative times $t-t'\lesssim 30\hbar/J$. For later times $t$, however, notable deviations are observable. Further, in panel (f), this range of qualitative agreement is mostly limited to times $t\lesssim 60\hbar/J$, i.e., relative times $t-t'\lesssim 10\hbar/J$. However, the oscillations have similar frequencies for later times $t$. Similar to the small chain with six sites, the polarization approximations display significantly larger amplitudes compared to the $GW$ approximation for almost all times $t$ for later $t'$ and stronger coupling. While in panel (e) the density response function for the $GW$ approximation remains within the interval between $-0.05/\hbar$ and $0.1/\hbar$, the polarization approximations lead to results for $\chi^\mathrm{R}$ that range from $-0.2/\hbar$ to $0.3/\hbar$. Similarly, in panel (f), we have $-0.2/\hbar\lesssim \chi^\mathrm{R}\lesssim 0.2/\hbar$ for the polarization approximations and $-0.1/\hbar\lesssim \chi^\mathrm{R}\lesssim 0.1/\hbar$ for the $GW$ approximation. \\
Overall, a comparison of the chains with six and thirty lattice sites shows that the agreement between the polarization approximations and the $GW$ approximation improves with increasing system size. In particular, it can be seen that PA and SPA-ME can be considered fully equivalent for larger systems. This has already been shown for the one-time PA and the stochastic polarization approximation (SPA) \cite{bonitz_pssb23} and is extended here to the two-time case.

\section{Discussion}\label{s:discussion}
In this paper, we have investigated the relation of the two-time quantum polarization approximation within the quantum fluctuations approach, as developed in Refs. \cite{schroedter_cmp_22,schroedter_23}, to the nonequilibrium Bethe--Salpeter equation. We examined the connection of the real-time components of the two-particle nonequilibrium Green function, as well as the exchange-correlation function, with two-time two-particle fluctuations. We applied the Hartree--Fock GKBA, within the Bethe--Salpeter approach, and derived a set of equations of motion for the two-time two-particle Green function in the $GW$ approximation, thus showing the close relation of the two approximations. Moreover, we showed the equivalence of the polarization approximation and the $GW$ approximation with additional exchange contributions in the weak coupling limit. The presented equations for $\mathcal{G}_2(t,t')$ constitute a generalization of the time-local version of the $GW$ approximation within the G1--G2 scheme \cite{schluenzen_prl_20,joost_prb_20,joost_prb_22} and allow for direct calculations of correlation functions of density and spin fluctuations as well as their corresponding structure factors. Furthermore, the equations describing the time-off-diagonal propagation have the same structure as the equations of motion on the diagonal. Overall, this shows that the numerical effort for the two-time approximations, i.e., $GW$ and PA, is comparable to the numerical cost of calculations of the $GW$ approximation in the G1--G2 scheme. This means that essentially any system that can be described in the G1--G2 scheme can also be treated using the two-time approximations.

However, computing these two-particle quantities, in general, has the drawback of having to store a rank-four tensor in the form of either $\mathcal{G}_2$ or $L$. With increasing system size, the computational demands of this approach, therefore, significantly increase. The present quantum fluctuations approach, on the other hand, proves to be more powerful as it allows for a combination with stochastic approaches such as the stochastic mean-field approximation \cite{ayik_plb_08,lacroix_prb14,lacroix_epj_14} that circumvent the direct computation of any rank-four tensors, in favor of an ensemble of single-particle  quantities (rank-two tensors). Within the context of the polarization approximation, this has led to the stochastic polarization approximation and its extension by means of the so-called multiple ensembles approach \cite{schroedter_cmp_22,schroedter_23}, which allows for the direct computation of correlation functions of fluctuations without having to store two-particle correlations or fluctuations. Most importantly, the main advantage of this approach lies in its applicability to correlated quantum systems that are far from equilibrium. This has been demonstrated in Ref.~\cite{schroedter_23} and further extended, in the present work, for Fermi--Hubbard chains. However, it must be noted that the special structure of the Fermi-Hubbard model leads to many simplifications in the equations. This is one of the reasons why it is important to consider other systems in future considerations, such as the homogeneous electron gas or other spatially uniform systems. This enables a more comprehensive characterization of the relationship between the polarization approximation and the $GW$ approximation and the interplay between fluctuations and correlations.

The theoretical analysis presented here further demonstrates that the connection of the quantum fluctuations approach and its approximations to the general theory of nonequilibrium Green functions proves to be particularly important for the future development of this theory. It allows for systematic extensions, such as the incorporation of stronger coupling or combination with embedding approaches \cite{balzer_prb_23}. Following these theoretical considerations and the extensive numerical tests for Fermi--Hubbard chains done here and in Ref.~\cite{schroedter_23}, further systematic analysis of the nonequilibrium dynamics of the correlation functions and its dependence on the coupling strength and system setup is of great interest. A particularly interesting excitation scenario, to be studied in the future, is given by short spatially monochromatic pulses. For uniform system, it was shown in Ref.~\cite{kwong_prl_00}, that such excitations lead to high quality dynamic structure factors. It is therefore to be expected that highly accurate results can also be obtained within the polarization approximation in similar scenarios.

\medskip
\textbf{Acknowledgements} \par
We thank J.-P. Joost, B. J. Wurst and C. Makait for many discussions on the quantum fluctuations approach. This work is supported by the German Science Foundation (DFG) via project BO1366-16.

%

\begin{thebibliography}{10}
\providecommand{\url}[1]{\texttt{#1}}
\providecommand{\urlprefix}{URL }

\bibitem{giuliani_vignale_2005}
G.~Giuliani, G.~Vignale,
\newblock \emph{Quantum Theory of the Electron Liquid},
\newblock Cambridge University Press, \textbf{2005}.

\bibitem{moreo_prb_93}
A.~Moreo,
\newblock \emph{Phys. Rev. B} \textbf{1993}, \emph{48} 3380.

\bibitem{lee_prb_03}
J.~D. Lee,
\newblock \emph{Phys. Rev. B} \textbf{2003}, \emph{67} 153108.

\bibitem{assaad_prb_06}
A.~Abendschein, F.~F. Assaad,
\newblock \emph{Phys. Rev. B} \textbf{2006}, \emph{73} 165119.

\bibitem{dornheim_physrep_18}
T.~Dornheim, S.~Groth, M.~Bonitz,
\newblock \emph{Phys. Rep.} \textbf{2018}, \emph{744} 1 .

\bibitem{dornheim_prl_18}
T.~Dornheim, S.~Groth, J.~Vorberger, M.~Bonitz,
\newblock \emph{Phys. Rev. Lett.} \textbf{2018}, \emph{121} 255001.

\bibitem{dornheim_pop_23}
T.~Dornheim, Z.~A. Moldabekov, K.~Ramakrishna, P.~Tolias, A.~Baczewski,
  D.~Kraus, T.~Preston, D.~Chapman, M.~B\"ohme, T.~Doeppner, F.~Graziani,
  M.~Bonitz, A.~Cangi, J.~Vorberger,
\newblock \emph{Physics of Plasmas} \textbf{2023}, \emph{30} 032705.

\bibitem{bonitz_pop_20}
M.~Bonitz, T.~Dornheim, Z.~A. Moldabekov, S.~Zhang, P.~Hamann, H.~Kählert,
  A.~Filinov, K.~Ramakrishna, J.~Vorberger,
\newblock \emph{Physics of Plasmas} \textbf{2020}, \emph{27}, 4 042710.

\bibitem{bonitz-etal.94pre}
M.~Bonitz, R.~Binder, D.~Scott, S.~Koch, D.~Kremp,
\newblock \emph{Phys. Rev. E} \textbf{1994}, \emph{49}.

\bibitem{dornheim_prl_20}
T.~Dornheim, J.~Vorberger, M.~Bonitz,
\newblock \emph{Phys. Rev. Lett.} \textbf{2020}, \emph{125} 085001.

\bibitem{dornheim_prr_21}
T.~Dornheim, M.~B\"ohme, Z.~Moldabekov, J.~Vorberger, M.~Bonitz,
\newblock \emph{Physical Review Research} \textbf{2021}, \emph{3} 033231.

\bibitem{gull_prb_19}
X.~Dong, X.~Chen, E.~Gull,
\newblock \emph{Phys. Rev. B} \textbf{2019}, \emph{100} 235107.

\bibitem{gull_prr_20}
S.~Li, E.~Gull,
\newblock \emph{Phys. Rev. Res.} \textbf{2020}, \emph{2} 013295.

\bibitem{pereira_prb_12}
R.~G. Pereira, K.~Penc, S.~R. White, P.~D. Sacramento, J.~M.~P. Carmelo,
\newblock \emph{Phys. Rev. B} \textbf{2012}, \emph{85} 165132.

\bibitem{kwong_prl_00}
N.-H. Kwong, M.~Bonitz,
\newblock \emph{Phys. Rev. Lett.} \textbf{2000}, \emph{{\bf 84}} 1768.

\bibitem{keldysh64}
L.~Keldysh,
\newblock \emph{Soviet Phys. JETP} \textbf{1965}, \emph{20} 1018,
  (Zh.~Eksp.~Teor.~Fiz.~\textbf{47}, 1515 (1964)).

\bibitem{stefanucci-book}
G.~Stefanucci, R.~van Leeuwen,
\newblock \emph{Nonequilibrium Many-Body Theory of Quantum Systems},
\newblock Cambridge University Press, \textbf{2013}.

\bibitem{balzer-book}
K.~Balzer, M.~Bonitz,
\newblock \emph{Nonequilibrium {G}reen's {F}unctions Approach to Inhomogeneous
  Systems},
\newblock Springer, Berlin Heidelberg, \textbf{2013}.

\bibitem{bonitz_pss_19_keldysh}
M.~Bonitz, A.-P. Jauho, M.~Sadovskii, S.~Tikhodeev,
\newblock \emph{physica status solidi (b)} \textbf{2019}, \emph{256}, 7
  1800600.

\bibitem{schluenzen_prb16}
N.~Schl{\"u}nzen, S.~Hermanns, M.~Bonitz, C.~Verdozzi,
\newblock \emph{Phys. Rev. B} \textbf{2016}, \emph{93} 035107.

\bibitem{schluenzen_prl_20}
N.~Schlünzen, J.-P. Joost, M.~Bonitz,
\newblock \emph{Phys. Rev. Lett.} \textbf{2020}, \emph{124}, 7 076601.

\bibitem{joost_prb_20}
J.-P. Joost, N.~Schl\"unzen, M.~Bonitz,
\newblock \emph{Phys. Rev. B} \textbf{2020}, \emph{101} 245101.

\bibitem{joost_prb_22}
J.-P. Joost, N.~Schl\"unzen, H.~Ohldag, M.~Bonitz, F.~Lackner, I.~Brezinova,
\newblock \emph{Physical Review B} \textbf{2022}, \emph{105} 165155.

\bibitem{donsa_prr_23}
S.~Donsa, F.~Lackner, J.~Burgdörfer, M.~Bonitz, B.~Kloss, A.~Rubio,
  I.~Brezinova,
\newblock \emph{Phys. Rev. Research} \textbf{2023}, \emph{5} 033022.

\bibitem{bonitz_pssb23}
M.~Bonitz, J.-P. Joost, C.~Makait, E.~Schroedter, K.~Balzer,
\newblock \emph{phys. stat. sol. (b)} \textbf{2023}.

\bibitem{schroedter_cmp_22}
E.~Schroedter, J.-P. Joost, M.~Bonitz,
\newblock \emph{Cond. Matt. Phys.} \textbf{2022}, \emph{25} 23401.

\bibitem{klimontovich_jetp_57}
Y.~Klimontovich,
\newblock \emph{JETP} \textbf{1957}, \emph{33} 982.

\bibitem{klimontovich_jetp_72}
Y.~Klimontovich,
\newblock \emph{JETP} \textbf{1972}, \emph{62} 1770.

\bibitem{klimontovich_1982}
Y.~L. Klimontovich,
\newblock \emph{Kinetic Theory of Nonideal Gases and Nonideal Plasmas},
\newblock Pergamon Press, \textbf{1982}.

\bibitem{ayik_plb_08}
S.~Ayik,
\newblock \emph{Phys. Lett. B} \textbf{2008}, \emph{658} 174.

\bibitem{lacroix_prb14}
D.~Lacroix, S.~Hermanns, C.~M. Hinz, M.~Bonitz,
\newblock \emph{Phys. Rev. B} \textbf{2014}, \emph{90} 125112.

\bibitem{lacroix_epj_14}
D.~Lacroix, S.~Ayik,
\newblock \emph{Eur. Phys. J. A} \textbf{2014}, \emph{50} 95.

\bibitem{filinov_prb_2}
V.~Filinov, P.~Thomas, I.~Varga, T.~Meier, M.~Bonitz, V.~Fortov, S.~W. Koch,
\newblock \emph{Phys. Rev. B} \textbf{2002}, \emph{{\bf 65}} 165124.

\bibitem{polkovnikov_ap_10}
A.~Polkovnikov,
\newblock \emph{Annals of Physics} \textbf{2010}, \emph{325}, 8 1790.

\bibitem{schroedter_23}
E.~Schroedter, B.~Wurst, J.-P. Joost, M.~Bonitz,
\newblock \emph{Phys. Rev. B} \textbf{2023}, \emph{108} 205109.

\bibitem{bonitz_qkt}
M.~Bonitz,
\newblock \emph{{Quantum Kinetic Theory}},
\newblock Teubner-Texte zur Physik. Springer, Cham, 2 edition, \textbf{2016}.

\bibitem{Lipavsky1986}
P.~Lipavsk\'y, V.~\ifmmode \check{S}\else \v{S}\fi{}pi\ifmmode~\check{c}\else
  \v{c}\fi{}ka, B.~Velick\'y,
\newblock \emph{Phys. Rev. B} \textbf{1986}, \emph{34} 6933.

\bibitem{schluenzen_jpcm_19}
N.~Schlünzen, S.~Hermanns, M.~Scharnke, M.~Bonitz,
\newblock \emph{Journal of Physics: Condensed Matter} \textbf{2020}, \emph{32}
  103001.

\bibitem{joost_phd_2022}
J.-P. Joost,
\newblock Dissertation, {Christian-Albrechts-Universität zu Kiel}, {Kiel},
  \textbf{2022}.

\bibitem{balzer_prb_23}
K.~Balzer, N.~Schl\"unzen, H.~Ohldag, J.-P. Joost, M.~Bonitz,
\newblock \emph{Phys. Rev. B} \textbf{2023}, \emph{107} 155141.

\end{thebibliography}

\end{document}